\documentclass[sigplan,10pt]{acmart}

\def\BibTeX{{\rm B\kern-.05em{\sc i\kern-.025em b}\kern-.08emT\kern-.1667em\lower.7ex\hbox{E}\kern-.125emX}}

\startPage{1}

\setcopyright{none}

\usepackage{li-stdpkgs}
\usepackage{li-typography}

\newcommand{\rv}{\textcolor{black}}

\usepackage{amsmath}
\usepackage{placeins}
\begin{document}

%
\title{A Parallel Sparse Tensor Benchmark Suite on CPUs and GPUs}

%
\author{Jiajia Li}
\email{jiajia.li@pnnl.gov}
\affiliation{%
  \institution{Pacific Northwest National Laboratory}
  \streetaddress{902 Battelle Blvd}
  \city{Richland}
  \state{WA}
  \country{USA}
  \postcode{99354}
}

\author{Mahesh Lakshminarasimhan}
\email{maheshl@cs.utah.edu}
\affiliation{%
  \institution{University of Utah}
  \city{Salt Lake City}
  \state{UT}
  \country{USA}
  \postcode{84112}
}

\author{Xiaolong Wu}
\email{xu1565@purdue.edu}
\affiliation{Purdue University%
  \institution{Electrical and Computer Engineering}
  \city{West Lafayette}
  \country{USA}
  \postcode{47907}
}

\author{Ang Li}
\email{ang.li@pnnl.gov}
\affiliation{%
  \institution{Pacific Northwest National Laboratory}
  \city{Richland}
  \state{WA}
  \country{USA}
  \postcode{99354}
}

\author{Catherine Olschanowsky}
\email{catherineolschan@boisestate.edu}
\affiliation{%
  \institution{Boise State University}
  \city{Boise}
  \state{ID}
  \country{USA}
  \postcode{83716}
}

\author{Kevin Barker}
\email{Kevin.Barker@pnnl.gov}
\affiliation{%
 \institution{Pacific Northwest National Laboratory}
  \city{Richland}
  \state{WA}
  \country{USA}
  \postcode{99354}
}

%
\renewcommand{\shortauthors}{Li et al.}

%
\begin{abstract}
Tensor computations present significant performance challenges that impact a wide spectrum of applications ranging from machine learning, healthcare analytics, social network analysis, data mining to quantum chemistry and signal processing. Efforts to improve the performance of tensor computations include exploring data layout, execution scheduling, and parallelism in common tensor kernels. This work presents a benchmark suite for arbitrary-order sparse tensor kernels using state-of-the-art tensor formats: coordinate (COO) and hierarchical coordinate (HiCOO) on CPUs and GPUs. It presents a set of reference tensor kernel implementations that are compatible with real-world tensors and power law tensors extended from synthetic graph generation techniques. We also propose Roofline performance models for these kernels to provide insights of computer platforms from sparse tensor view.

\end{abstract}

\maketitle

\section{Introduction} \label{sec:intro}

Tensors, multi-dimensional arrays that are often sparse, are utilized by a large number of critical applications that span a range of domain areas. 
These include quantum chemistry, healthcare analytics, social network analysis, data mining, signal processing, machine learning, and more.
Operations on sparse tensors tend to dominate the execution-time of these applications.
Understanding the performance characteristics of different implementation approaches is of paramount importance.
This paper presents a benchmark suite specifically for that purpose. 
The suite provides implementations of common tensor kernels using state-of-the-art sparse tensor data structures and a variety of real and synthetic sparse tensors as its input dataset.

Given the heterogeneity in available hardware resources for high performance computing (HPC), it is non-trivial to answer questions about the potential for sparse tensor algorithms to be efficiently ported to various hardware.
The difficulty of planning for the irregular parallelism that results from operating on sparse data structures is compounded by the availability of Graphics Processing Units (GPUs), vectorizing units, Field Programmable Gate Arrays (FPGAs), and potentially Tensor Processing Units (TPUs). 
A set of important tensor kernels with associated implementations eases the exploration of this space.

Optimizing the performance of tensor applications is challenging due to several application characteristics, named in~\cite{Kolda:2009:survey,Cichocki:2014:survey,li2019pasta,Li:2018:thesis} and briefly outlined here for completeness: \emph{the curse of dimensionality, mode orientation, tensor transformation, irregularity, and arbitrary tensor orders (or dimensions)}.

Tensors are, by definition, multidimensional. 
The \emph{curse of dimensionality} manifests itself as large computational and storage overheads required to accommodate the exponential growth of elements that occurs in some operations.
For instance, a Kronecker product results in exponential expansion of space requirements.
Compounding this issue is the increased interest in applications involving a large number of dimensions~\cite{DeLathauwer:2017:tensorize,Lebedev:2014:cnn-cp,Novikov:2015:tnn,Li:2017:adatm,Kaya:2018:dimtree-sparse-cp}.
The data structures supporting sparse tensors and the required tensor operations are often mode specific, where each dimension of a tensor is referred to as a mode.
Different data structures supporting sparse tensors favor iterating over specific modes, \emph{mode orientation}.
There is a tradeoff that must be made between space requirements and enjoying good performance in multiple representations of various mode sequences.
\emph{Tensor transformation} is traditionally used to implement tensor operations by casting them as a set of matrix operations and utilizing highly tuned linear algebra libraries. 
However, the transformation process brings non-trivial overhead to the execution of a tensor operation.
Mitigating this cost has become attractive for researchers in tensor linear algebra and their applications~\cite{Li:2015:intensli,DiNapoli:2014:tc,Matthews:2016:tc,Springer:2017:HPTT,Li:2016:spttm}.
\emph{Irregularity} in memory access patterns and in tensor shape makes poor use of memory subsystems and complicates code, especially for sparse data. 
Optimizations are typically best suited for a specific dimensionality, such as third-order, but most tensor operations are required to handle \emph{arbitrary tensor orders}.

Beyond these, challenges associated with all benchmarks also apply, which include \emph{completeness, diversity, extendibility, reproducibility, and comparability} across implementations.
Comparisons across research groups are improved by using a standard set of kernels and inputs.
Using that set as a starting point, optimizations can be applied and effectively compared.

Our benchmark suite consists of a set of reference implementations from various tensor applications, each of which show different computational behavior. 
Much like\\ two-dimensional sparse matrices, the \emph{data layout}, or the data structure used to hold a sparse tensor, has a significant impact on performance and storage~\cite{Li:2013:smat,sedaghati2015spmv}. 
It also has a significant impact on how the control flow for a given operation must be executed and its memory footprints.
We implement two sparse tensor formats: the most popular and mode-generic coordinate (COO) format and a newly proposed, more compressed hierarchical coordinate (HiCOO) format~\cite{Li:2018:hicoo} to represent general, arbitrary sparse tensors.
Beyond the implementation diversity, platform and workload (or input) diversity is also critical to gain insights from a benchmark suite.
We implement the same set of tensor kernels on CPUs and GPUs to provide a good understanding for users.
Different inputs of an algorithm usually obtain different performance due to their diverse data sizes and patterns.
This phenomenon is more obvious for sparse problems because their algorithm behavior largely depends on the features of data.
Besides evaluating limited and hard-to-obtain real-world tensors, mimicking some application characters to generate more datasets is valuable for benchmarking.
We create power law tensors extended from synthetic graph generation techniques.
Our benchmark suite can easily adopt new sparse tensor kernels. We use floating point operations per second (FLOPS) to compare between kernels and platforms.

The contributions of this work include:
\begin{compactitem}
  \item reference implementations for five tensor kernels, \TEW, \TS, \TTV, \TTM, and \MTTKRP, in COO and HiCOO formats for CPUs and GPUs; (\Cref{sec:kernels,sec:impl})
  \item application of HiCOO to more tensor operations and an extension of it to more flexible variations; (\Cref{sec:impl})
  \item synthetic tensor generation based on Kronecker and power law generators; (\Cref{sec:dataset})
  \item Roofline performance models for two Intel CPU and two NVIDIA GPU platforms to analyze the tensor kernels; and
  \item insights gained from thorough experiments and analysis of the performance. (\Cref{sec:exp})
\end{compactitem}

    
    


\section{Tensor Benchmarks} \label{sec:kernels}
Tensors are increasingly employed in computations across a spectrum of application areas. 
This benchmark suite represents a set of basic operations chosen by examining a range of composite operations commonly used by these applications.
\rv{The following text provides the definition of each operation, the motivation for its inclusion, and its applications.}

\rv{
Notationally, we represent tensors as calligraphic capital letters, e.g., $\T{X} \in \R^{I \times J \times K}$;
matrices by boldface capital letters, e.g., $\M{U} \in \R^{I \times J}$;
vectors by boldface lowercase letters, e.g., $\V{x} \in \R^I$;
and scalars by lowercase letters, such as $x_{ijk}$ for the $(i,j,k)$-element of a third-order tensor $\T{X}$.
A \emph{slice} is a two-dimensional cross-section of a tensor, obtained by fixing all indices but two, e.g., $\M{S}_{::k} = \T{X}(:,:,k)$. 
A \emph{fiber} is a vector extracted from a tensor along a certain mode, selected by fixing all indices but one, e.g., $\V{f}_{:jk} = \T{X}(:,j,k)$. 
}

\subsection{Tensor Element-Wise Operations}
\label{sec:tew}

Tensor element-wise (\TEW) operations include addition, subtraction, multiplication, and division, that are applied to every corresponding pair of elements from two tensor objects if they have the same order and shape (i.e., dimension sizes).
For example, element-wise tensor addition of $\T{X}, \T{Y} \in \R^{I_1 \times \dots \times I_N}$ is
\begin{equation}
\begin{split}
\T{Z} &= \T{X} + \T{Y}\\
\end{split}
\label{eq:tew}
\end{equation}
Similarly for element-wise tensor subtraction $\T{Z} = \T{X} - \T{Y}$, multiplication $\T{Z} = \T{X} \circ \T{Y}$, and division $\T{Z} = \T{X} \varoslash \T{Y}$. 

This operation is trivially implemented when the tensors having exactly the same non-zero pattern. 
However, the more general case requires iterating over both tensors and matching elements as the execution proceeds.
When \T{X} and \T{Y} have different patterns, predicting the storage required for \T{Z} is an additional challenge.

\subsection{Tensor-Scalar Operations}
\label{sec:ts}
A Tensor-Scalar (\TS) operation is between the non-zero values of a tensor and a single scalar.
Operations also include addition (\TSA), subtraction (\TSS), multiplication (\TSM), and division (\TSD).
For example, tensor-scalar multiplication of $\T{X} \in \R^{I_1 \times \dots \times I_N}$ is

\begin{equation}
	\begin{split}
	\T{Y} &= \T{X} \times s	\\
	\end{split}
\label{eq:ts}
\end{equation} 
This benchmark suite implements only \TSA and \TSM, which are sufficient to suppor them all. 

\rv{\TEW and \TS are commonly used in machine learning, quantum chemistry, and more.
Tensor convolution operation in convolutional neural network (CNN) is a combination of \TEW and \TSM~\cite{ji20123d}; space mapping in quantum chemistry also involves these two.
}
\rv{\TEW and \TS are simple tensor operations and can be implemented along with other tensor operations.
We consider them separately in this benchmark suite because of their different computational behavior (shown in \Cref{tab:analysis}).
}
 
\subsection{Tensor-Times-Vector Product}
\label{sec:ttv}
The Tensor-Times-Vector (\TTV) in mode $n$ is the multiplication of a tensor $\T{X} \in \R^{I_1 \times \dots \times I_n \times \dots \times I_N}$ with a vector $\V{V} \in \R^{I_n}$, along mode $n$.
\begin{equation}
	\begin{split}
	\T{Y} = \T{X} \times_n \V{V}\\	
	y_{i_1 \cdots i_{n-1} i_{n+1} \cdots i_N} &= \sum_{i_n=1}^{I_n} x_{i_1 \cdots i_{n-1} i_n i_{n+1} \cdots i_N} v_{i_n}
	\end{split}
\label{eq:ttv}
\end{equation}
This results in a $I_1 \times \dots \times I_{n-1} \times I_{n+1} \times \dots \times I_N$ tensor which has one less dimension.

\rv{\TTV is a critical computational kernel of the tensor power method~\cite{Anandkumar:2014:survey,Yu:2018:longterm}, an approach for orthogonal tensor decomposition, that decomposes a symmetric tensor into a collection of orthogonal vectors with corresponding weights. 
The tensor power method is used in machine learning and signal processing applications.
}

\subsection{Tensor-Times-Matrix Product}
\label{sec:ttm}
The Tensor-Times-Matrix (\TTM) in mode $n$, also known as the $n$-mode product, is the multiplication of a tensor $\T{X} \in \R^{I_1 \times \dots \times I_n \times \dots \times I_N}$ with a matrix $\M{U} \in \R^{I_n \times R}$, along mode $n$, and is denoted by $\T{Y} = \T{X} \times_n \M{U}$.
\footnote{Our convention for the dimensions of \M{U} differs from that of Kolda and Bader's definition~\cite{Kolda:2009:survey}. In particular, we transpose the matrix modes \M{U}, which leads to a more efficient \TTM under the row-major storage convention of the C language.}
This results in a $I_1 \times \dots \times I_{n-1} \times R \times I_{n+1} \times \dots \times I_N$ tensor, and its operation is defined as
\begin{equation}
y_{i_1 \cdots i_{n-1} r i_{n+1} \cdots i_N} = \sum_{i_n=1}^{I_n} x_{i_1 \cdots i_{n-1} i_n i_{n+1} \cdots i_N} u_{i_n r}.
\label{eq:ttm}
\end{equation}
\TTM is a special case of tensor contraction\rv{, a multiplication between two tensors in common mode(s).}
We consider \TTM specifically because: 1) it is more commonly used in tensor decompositions, such as the Tucker decomposition, for \rv{a variety of applications, including (social network, electrical grid) data analytics, numerical simulation, machine learning, recommendation systems, personalized web search, etc.~\cite{Kolda:2009:survey,Cichocki:2016:survey,Anandkumar:2014:survey,Sidiropoulos:2017:survey}}; \rv{2) the behavior of tensor contraction largely depends on which mode(s) to be contracted on; this creates difficulties for benchmarking}.
Also, note that $R$ is typically much smaller than $I_n$ in low-rank decompositions, and typically $R < 100$.

\subsection{Matriced Tensor-Times-Khatri-Rao Product}
\label{sec:mttkrp}
\MTTKRP, matricized tensor times Khatri-Rao product, is a matricized tensor times the Khatri-Rao product of matrices.
For an $N$th-order tensor \T{X} and given matrices $\M{U}^{(1)}, \ldots, \M{U}^{(N)}$, the mode-$n$ \MTTKRP is
\begin{equation}
\small
\begin{split}
\tilde{\M{U}}^{(n)} 
	&= \M{X}_{(n)} \left( \M{U}^{(N)} \odot \cdots \odot \M{U}^{(n+1)} \odot \M{U}^{(n-1)} \odot \cdots \odot \M{U}^{(1)} \right),
\label{eq:mttkrp}
\end{split}
\end{equation}
where $\M{X}_{(n)}$ is the mode-$n$ matricization of tensor $\T{X}$, $\odot$ is the Khatri-Rao product.
The Khatri-Rao product is a ``matching column-wise'' Kronecker product between two matrices.
Given matrices $\M{A} \in \R^{I \times R}$ and $\M{B} \in \R^{J \times R}$, their Khatri-Rao product is denoted by $\M{C} = \M{A} \odot \M{B}$,
\begin{equation}
\small
{
\M{C} = \M{A} \odot \M{B} = \left[ \V{a}_1 \circ \V{b}_1, \V{a}_2 \circ \V{b}_2, \dots, \V{a}_R \circ \V{b}_R  \right],
}
\end{equation}
where $\M{C} \in \R^{(IJ) \times R}$, $\V{a}_r$ and $\V{b}_r$, $r=1,\dots,R$ are columns of $\M{A}$ and $\M{B}$, $\circ$ is the outer product of vectors, a special case of Kronecker product.
However, the Khatri-Rao and Kronecker products typically require redundant computation or extra storage to hold matrix operands, so in practice, these operations tend to be not implemented directly but rather integrated into tensor operations.

\rv{\MTTKRP is the most computational expensive kernel in CANDECOMP/PARAFAC decomposition (CPD), another popular tensor decomposition.
CPD also has a wide application in (healthcare, social network, brain signal, electrical grid) data analytics, machine learning, recommendation systems, signal processing, personalized web search, quantum chemistry, and other domains~\cite{Kolda:2009:survey,Cichocki:2016:survey,Anandkumar:2014:survey,Sidiropoulos:2017:survey}.
}

\rv{
Because of the varying computational behavior (shown in \Cref{tab:analysis}) of the above operations, we integrate them as a benchmark suite to help evaluate underlying hardware. 
}




\section{Tensor Formats and Kernel Implementations} \label{sec:impl}

Much like sparse matrices, sparse tensors are expressed using different formats.
The best choice of format depends on the sparsity pattern of a tensor, operations applied, and the time required to translate between them. 
The common default format for sparse tensors is coordinate (COO) format. 
New formats have been developed including compressed sparse fiber (CSF)~\cite{Smith:2015:splatt}, balanced CSF (BCSF)~\cite{Nisa:2019:BCSF}, flagged COO (F-COO)~\cite{Liu:2017:fcoo}, and hierarchical coordinate (HiCOO)~\cite{Li:2018:hicoo} for general sparse tensors, and mode-generic and -specific formats for structured sparse tensors~\cite{Baskaran:2012:sparse-tensor}.
Our benchmark suite currently supports COO and HiCOO for general sparse tensors and their variants for semi-sparse tensors with dense dimension(s). 

We chose COO and HiCOO formats because of their \emph{mode generic} property, described in~\cite{Li:2018:hicoo}. 
When using a mode generic format only one tensor representation is needed for all tensor computations, even in different modes. This is commonly required by tensor methods (e.g., CPD and Tucker decompositions). 
COO is the most popular format used in many tensor libraries, e.g., Tensor Toolbox~\cite{Bader:2017:tensortoolbox-pak}, Tensorlab~\cite{Vervliet:2016:tensorlab-pak}, TACO~\cite{chou:2018:formats,kjolstad:2017:taco}, and ParTI~\cite{Li:2016:parti-pak}.
HiCOO, a newly proposed format, obtains good compression and state-of-the-art tuned performance~\cite{Li:2018:hicoo}.
Other formats especially CSF will be considered for our benchmark suite in the near future.
This section overviews our supported formats and their corresponding parallel implementations for tensor kernels.

We keep the implementations simple yet effective; the benchmark represents a general case where the primary computation is not obfuscated by optimization attempts.
We use more preprocessing to trade for less kernel computation, which keeps the benchmark more efficient compared to the pillar libraries, e.g., Tensor Toolbox~\cite{Bader:2017:tensortoolbox-pak} and TensorLab~\cite{Vervliet:2016:tensorlab-pak}.
Besides, our implementations directly operate on sparse tensor elements to avoid the tensor-matrix transformations.
This suite provides a performance baseline and a starting point for new optimization strategies.
It is easy to adopt new implementations, operations, and formats from users and to be adapted in a communication scheme.


\subsection{Coordinate Format (COO)}
\begin{figure}[hbt]
  \includegraphics[width=.4\columnwidth]{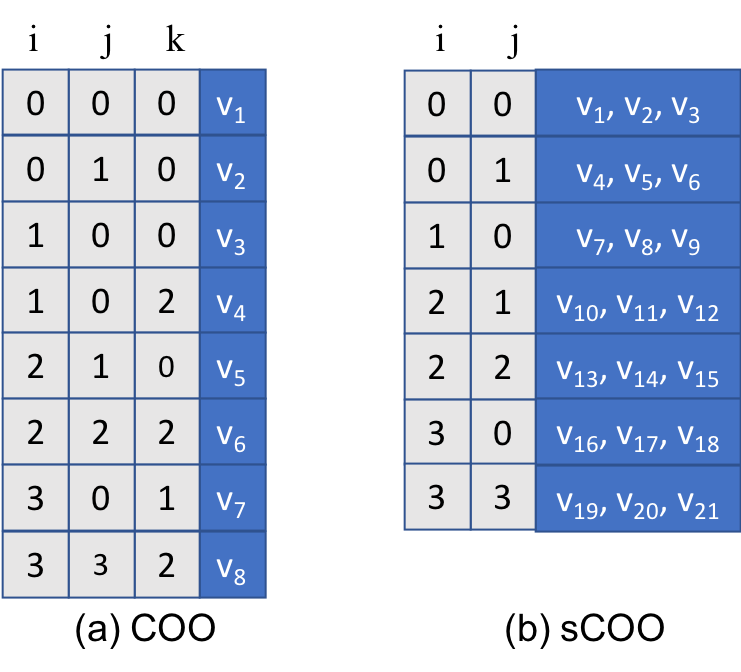}
  \caption{COO format for a general sparse tensor and sCOO format~\cite{Li:2016:spttm} for a semi-sparse tensor.}
  \label{fig:coo}
\end{figure}

Coordinate format is commonly used to store sparse matrices and tensors.
It does not require or guarantee any specific ordering of the data.
The data values are stored in a one-dimensional array, no matter how many dimensions are represented in the data.
For each dimension an additional index array is added that indicates the position of the value in that dimension. 
Figure~\ref{fig:coo}(a) gives an example that a general third-order sparse tensor requires three index arrays.
The storage space of an $N$th-order COO tensor $\T{X} \in \R^{I_1 \times \dots \times I_N}$ with $\nnz$ non-zeros is $4 (N+1) M$ bytes consisting of 32-bit indices and single-precision floating-point values.

We also describe a variant of COO format (semi-sparse COO, sCOO) for a semi-sparse sparse tensor with dense modes~\cite{Li:2016:spttm,Baskaran:2012:sparse-tensor}, which will be used in \TTM. A dense mode means the fibers on it are all dense.
sCOO stores the dense mode(s) as dense array(s) and the rest modes are kept the same as in COO format, as shown by another example tensor in~\Cref{fig:coo}(b), \rv{where the mode k is dense}. 

\subsection{Parallel Implementations based on COO}

\Cref{tab:analysis} presents the operational intensity of each kernel using a cubical third-order tensor, while all the implementations in the benchmark suite support arbitrary tensor orders.
Operational intensity (OI) is the ratio of bytes required per floating point operation for a given computation.
For all operations except \MTTKRP, we have a pre-processing stage to allocate output tensor space along with their indices.

The implementations of \TEW and \TS follow directly from their Equations, (\ref{eq:tew}) and (\ref{eq:ts}), that is one loop over all non-zero values to do the corresponding computation.
In the pre-precessing stage, we allocate and set indices for the output tensors due to their easy-to-predict non-zero pattern. \footnote{For \TEW with two input tensors have different non-zero patterns, we support it in the benchmark suite but not analyze them for performance perspective.}
\TEW and \TS have the smallest operational intensity: $1/12$ and $1/8$.
We will use \TTV algorithm as a representative to explain the similar \TTM algorithms as well; \TTM algorithms can be found in \cite{Li:2016:spttm,Ma:2018:sptucker-gpu}, and \MTTKRP algorithms can be easily found in  literatures and software~\cite{Li:2018:hicoo,Bader:2017:tensortoolbox-pak,Vervliet:2016:tensorlab-pak,Li:2016:parti-pak}.


\begin{table}[ht!]
\centering
\footnotesize
{
\caption{The analysis of kernel algorithms for third-order cubical tensors ($\T{X} \in \R^{I \times I \times I}$). We consider all input tensors with \nnz non-zero entries and $\nnz_F$ fibers, $I \ll \nfibs \ll \nnz$. The indices use $32$ bits, and values are single-precision floating-point numbers with $32$ bits as well.}
\label{tab:analysis}
\begin{tabular}{r|c|cc|c}
  \toprule
  
  \multirow{2}{*}{Kernels} & 
  Work & 
  \multicolumn{2}{c|}{Memory Access (\#Bytes)} & 
  OI \\

  &
  (\#Flops) &  
  COO &
  HiCOO &
   \\ \hline

  \TEW & 
  $\nnz $ & 
  $12 \nnz$ &
  $12 \nnz$ &
  $1/12$ \\\hline 

  \TS & 
  $\nnz$ & 
  $8 \nnz$ &
  $8 \nnz$ &
  $1/8$ \\ \hline

  \multirow{2}{*}{\TTV} & 
  \multirow{2}{*}{$2 \nnz$} & 
  \multirow{2}{*}{$12 \nnz + 12 \nfibs$} &
  \multirow{2}{*}{$12 \nnz + 12 \nfibs$} &
  \multirow{2}{*}{$\sim 1/6$} \\

  & 
  &
  &
  &
  \\ \hline

  \multirow{2}{*}{\TTM} & 
  \multirow{2}{*}{$2 \nnz R$} & 
  $4 \nnz R + 4 \nfibs R$ & 
  $4 \nnz R + 4 \nfibs R$ & 
  \multirow{2}{*}{$\sim 1/2$} \\

  & 
  & 
  $+ 8 \nnz + 8 \nfibs$ &
  $+ 8 \nnz + 8 \nfibs$ &
  \\ \hline

  \multirow{2}{*}{\MTTKRP} & 
  \multirow{2}{*}{$3 \nnz R$} & 
  \multirow{2}{*}{$12 \nnz R + 16 \nnz$} &
  $12 R min\{n_b \nnzb, \nnz\}$ &
  \multirow{2}{*}{$\sim 1/4$} \\

  & 
  & 
  &
  $+ 7 \nnz + 20 n_b$ &
  \\

  \bottomrule
\end{tabular}
\\ $^*$ \footnotesize{We consider only one-level Cache with the minimum cache size to satisfy the data reuse in algorithms.}
}
\end{table}

\subsubsection{Multicore CPU}
We use a \emph{sparse-dense property} for a sparse tensor times a dense vector/matrix (\TTV and \TTM), introduced by Li et al.~\cite{Li:2016:spttm}.
That is, if the computation is between a sparse mode of a tensor and a dense vector from the vector itself or a matrix, this mode will become dense in the output; and the other modes keep the same non-zero distribution (or sparsity) with the original modes of the input tensor.
This property makes pre-allocating space for the outputs of \TTV and \TTM possible, with the help of the sCOO format for semi-sparse tensors.
Introducing this property is good for parallelization by avoiding output data races and dynamic memory allocation, especially useful for GPU implementations.

\begin{algorithm}[h]
\footnotesize
\caption{COO-\TTV-OMP algorithm.}
\begin{algorithmic}[1]
  \Require
       A third-order sparse tensor $\T{X} \in \R^{I \times J \times K}$ with \nnz non-zeros in COO format, dense vector $\M{V} \in \R^{K}$, and an integer $n$ ($=3$);
  \Ensure
       Sparse tensor $\T{Y} \in \R^{I \times J}$ in COO format;

  \Statex \Comment{$\T{Y} = \T{X} \times_n \V{V}$} 
  \State Pre-process to obtain $\nfibs$, the number of mode-n fibers of \T{X} and $\fptr$, the beginning of each \T{X}'s mode-n fiber, size $\nfibs$.
  \State Pre-allocate \T{Y} space with $\nfibs$ non-zeros and their indices; 
  \Statex

  \PARFOR{$f=1, \dots, \nfibs$}
    \State $f_X = \fptr(f)$
    \State $\inds^1_Y(f) = \inds^1_X(f_X)$
    \State $\inds^2_Y(f) = \inds^2_X(f_X)$
    \State $v = \val_Y(f)$

    \For{$m=f_X, \dots, \fptr(f+1)-1$}
      \State $k = \inds^3_X(m)$
      \State $v += \val_X(m) \times u(k)$
    \EndFor
  \ENDPARFOR

  \State \textbf{Return} \T{Y};
\end{algorithmic}
\label{alg:ttv}
\end{algorithm}

\textbf{COO-\TTV-OMP} is illustrated in \Cref{alg:ttv}, firstly proposed in this work. 
We first pre-process the input tensor \T{X} to record the locations of mode-$n$ fibers.
According to the sparse-dense property, the output \T{Y} is pre-allocated with \nfibs non-zeros and its indices $i,j$ the same as in \T{X}.
The storage is consists of $16 \nnz$ for \T{X}, $4I$ for \V{v}, and $12 \nfibs$ for \T{Y} \footnote{which is actually a matrix for a third-order \TTV.}. 
The number of floating-point operations (\#Flops) is $2\nnz$.
The memory access in \Cref{tab:analysis} counts $4\nnz$ bytes for \V{V} because of its irregular and unpredictable memory access introduced by index-$k$.
Data reuse of \V{V} could happen if its access has or gains a good localized pattern naturally or from reordering techniques~\cite{Smith:2015:splatt,Li:2019:reorder}, similarly for the matrices in \TTM and \MTTKRP.
Thus, their performance could be improved due to reductions in memory pressure.
The operational intensity is approximately $1/6$ by ignoring less significant terms.
COO-\TTM-OMP is similar to COO-\TTV-OMP with the output as a semi-sparse tensor stored in sCOO format.
They are both parallelized for independent fibers, but work imbalance may exist because of different fiber lengths of sparse tensor \T{X}.



COO-\MTTKRP is widely used in Tensor Toolbox~\cite{Bader:2017:tensortoolbox-pak}, Tensorlab~\cite{Vervliet:2016:tensorlab-pak}; and \textbf{COO-\MTTKRP-OMP} is implemented in ParTI library~\cite{Li:2016:parti-pak}.
Each row of $\tilde{\M{A}}$ is updated by scaling the dot product of two rows of matrices \M{B} and \M{C} by the non-zero value of \T{X}.
Its operational intensity is approximately $1/4$ again by ignoring less expensive terms.
COO-\MTTKRP-OMP is parallelized by non-zeros, but with atomic operations to protect output matrix.
The data race may influence its performance differently depending on non-zero distributions of an input tensor.

\subsubsection{GPU}
COO-\TEW-GPU,-\TS-GPU, and -\TTV-GPU use one-dimensional CUDA grids of one-dimensional thread blocks to parallelize non-zeros and fibers respectively.
For example, \nnz non-zeros are assigned to $\nnz/256$ thread blocks with 256 threads for each.
Again, due to the potential unbalanced fiber lengths, COO-\TTV-GPU could suffer more performance drop.
While COO-\TTM-GPU and -\MTTKRP-GPU use one-dimensional grids of two-dimensional thread blocks to parallelize the dense matrices, both of them are implemented in ParTI library~\cite{Li:2016:parti-pak}.
In COO-\TTM-GPU algorithm, the x-dimension of thread blocks are used to represent matrix columns for GPU memory coalescing, while y-dimension represents non-zeros. (Refer to details in~\cite{Ma:2018:sptucker-gpu}.)
Be aware that the load imbalance still exists for COO-\TTV and COO-\TTM, and also data race for COO-\MTTKRP.

\subsection{Hierarchical Coordinate (HiCOO)}
\begin{figure}[hbt]
  \includegraphics[width=0.8\columnwidth]{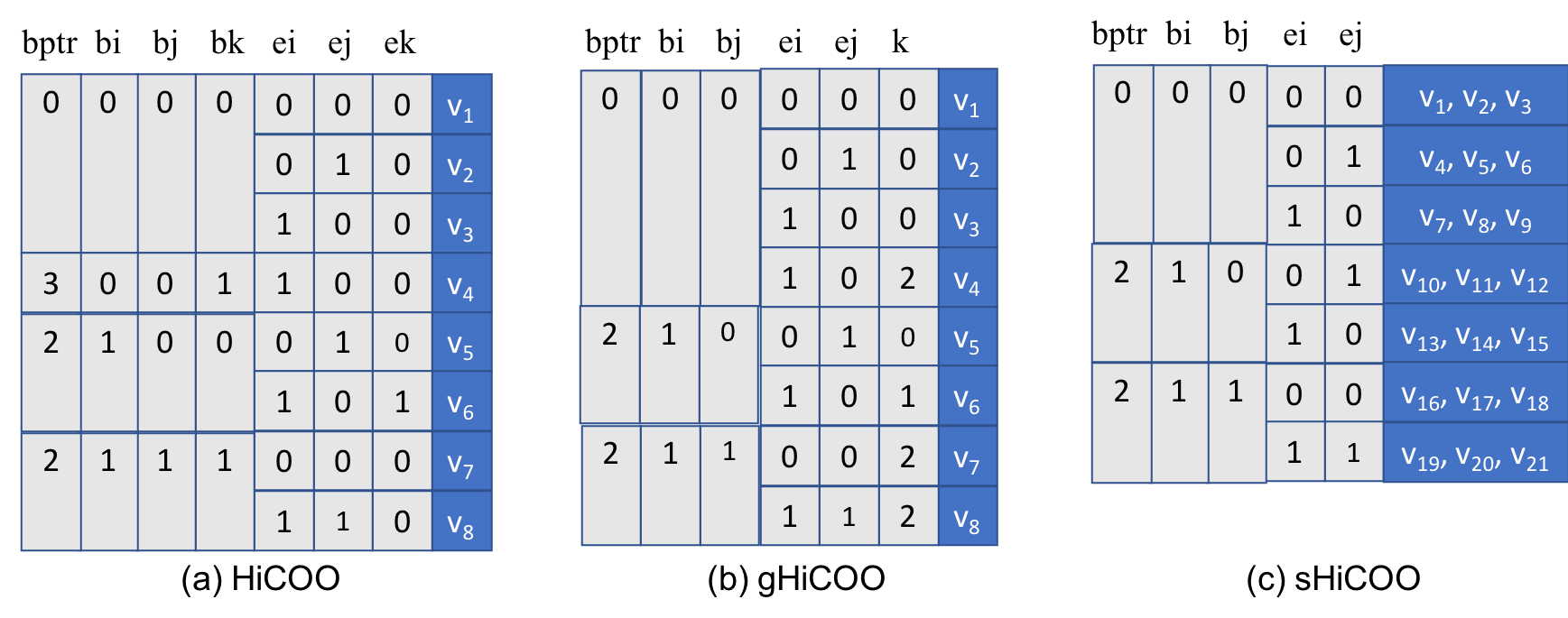}
  \caption{HiCOO, gHiCOO formats for general sparse tensors and sHiCOO for semi-sparse tensors.}
  \label{fig:hicoo}
\end{figure}

Hierarchical Coordinate (HiCOO)~\cite{Li:2018:hicoo} is derived from COO format but further compresses tensor indices in units of sparse blocks with a pre-specified block size $B$.
It represents tensor indices using two-level \emph{block} and \emph{element} indices, with element indices stored in only $8$-bit.
An extra block pointer array $bptr$ is needed to store the starting locations of every fiber. 
\Cref{fig:hicoo}(a) shows HiCOO representation for the same tensor in \Cref{fig:coo}(a) in $2 \times 2 \times 2$ blocks.
The same with COO format, HiCOO does not assume any mode order, and only one representation of a sparse tensor is enough for all its computations.
While HiCOO saves the tensor storage from two aspects: 1) shorter bit-length for element indices; 2) shortened array length for block indices.
Readers can refer to more details in the paper~\cite{Li:2018:hicoo}.

In this work we introduce two variants based on HiCOO format: gHiCOO and sHiCOO.
gHiCOO is a generalization of HiCOO format for a general sparse tensor (\Cref{fig:hicoo}(b)) and sHiCOO is for semi-sparse tensors with dense mode(s) (\Cref{fig:hicoo}(c)).
Concluded by the prior work~\cite{Li:2018:hicoo}, HiCOO could not be beneficial for hyper-sparse tensors where most tensor blocks only consist of one or few non-zeros.
To conquer this problem, we propose gHiCOO where we can pick which modes to be compressed in units of blocks for HiCOO and which stay in COO format. 
\Cref{fig:hicoo}(b) chooses to compress modes i and j, leaving mode k in the same index array with \Cref{fig:coo}(a).
gHiCOO also provide convenience to implement tensor operations where not all modes are needed during computation, such as \TTV and \TTM.
sHiCOO is similar to sCOO but in HiCOO format.
\Cref{fig:hicoo}(c) uses sHiCOO to compress the same semi-sparse tensor in \Cref{fig:coo}(b) with dense mode k.
Our format variants could be useful in tensor methods and benchmarking for more efficient space and computation.

\subsection{Parallel Implementations based on HiCOO}
HiCOO parallel implementations are all firstly proposed here except \MTTKRP on CPUs~\cite{Li:2018:hicoo}.
Advanced techniques such as privatization, lock-avoiding parallel strategies, advanced scheduling~\cite{Li:2018:hicoo} are not adopted as the purpose of this suite is to act as a reference implementation for the community and also to avoid complicated tuning parameters.

\subsubsection{Multicore CPU}
Since the pre-processing step deals with allocating space and setting indices for the output tensor in HiCOO rather than COO format, the value computation of HiCOO-\TEW-OMP and HiCOO-\TS-OMP is the same with COO-\TEW-OMP and COO-\TS-OMP respectively. 


\textbf{HiCOO-\TTV-OMP} and \textbf{HiCOO-\TTM-OMP} also pre-allocate the output tensors according to the sparse-dense property.
We use gHiCOO format to represent the input tensor \T{X} by leaving the mode doing the product uncompressed.
Therefore, \TTV and \TTM can bypass the blocking nature of HiCOO and be performed without data race between blocks.
Then the same computation will be implemented for HiCOO-\TTV-OMP and HiCOO-\TTM-OMP as in their COO counterparts, but pre-allocating the output tensors and their indices in HiCOO or sHiCOO format respectively.

\begin{algorithm}[h]
\footnotesize
\caption{HiCOO-\MTTKRP-OMP algorithm in mode-1~\cite{Li:2018:hicoo}.}
\begin{algorithmic}[1]
\footnotesize
  \Require
       A sparse tensor $\T{X} \in \R^{I \times J \times K}$ with \nnz non-zeros in HiCOO format, dense matrices $\M{B} \in \R^{J \times R}, \M{C} \in \R^{K \times R}$, block size $B$;
  \Ensure
       Updated dense matrix $\tilde{\M{A}} \in \R^{I \times R}$;

  \Statex \Comment{$\tilde{\M{A}} \leftarrow \T{X}_{(1)} (\M{C} \odot \M{B})$} 

  \PARFOR{$b=1, \dots, n_b$} 
  	\State $bi = \binds^1(b)$, $bj = \binds^2(b)$, $bk = \binds^3(b)$;
    \State $\M{A}_b = \M{A} + bi \cdot B \cdot R$; $\M{B}_b = \M{B} + bj \cdot B \cdot R$; $\M{C}_b = \M{C} + bk \cdot B \cdot R$;
    \For{$x=bptr(b), \dots, bptr(b+1) - 1$} \Comment{entry $x$}
      \State $ei = \einds^1(x)$, $ej = \einds^2(x)$, $ek = \einds^3(x)$
      \State $value = \val(x)$
      \For{$r=1, \dots, R$}
        \State  $\tilde{A}_b(ei,r) += value \cdot C_b(ek,r) \cdot B_b(ej,r)$
      \EndFor
    \EndFor
  \ENDPARFOR  
  \State \textbf{Return} $\tilde{\M{A}}$;
\end{algorithmic}
\label{alg:mttkrp-hicoo}
\end{algorithm}

\textbf{HiCOO-\MTTKRP-OMP} is more complicated because indices of all tensor modes are used in this computation, different from \TTV and \TTM where the indices from only one mode are needed.
We first block matrices \M{A}, \M{B}, and \M{C} as $\M{A}_b$, $\M{B}_b$, and $\M{C}_b$ to be reused for the non-zeros inside a tensor block.
For each block, we update the values of a block of output matrix $\tilde{\M{A}}$ using corresponding blocks $\M{B}_b$, and $\M{C}_b$.
In this way, we do not need to compute the actual indices $i, j, k$ out, and data locality is increased due to blocking and Morton order sorting when constructing a HiCOO representation~\cite{Li:2018:hicoo}.
Different from COO-\MTTKRP-OMP, HiCOO-\MTTKRP-OMP parallelize tensor blocks rather than all non-zeros.

The analysis of HiCOO algorithms are also listed in \Cref{tab:analysis}.
Since HiCOO-\TEW, -\TS, -\TTV, -\TTM all have the same value computation step with COO counterparts, so they have the same behavior except for the storage space, where HiCOO is usually beneficial.
However, from our experiments, we still observe some benefits of HiCOO affected by the pre-processing stage.
However, HiCOO-\MTTKRP has smaller memory access than COO-\MTTKRP due to its blocked feature.

\subsubsection{GPU} 
HiCOO-GPU implementations are also the same with COO ones except HiCOO-\MTTKRP-GPU.
This unoptimized HiCOO-\MTTKRP-GPU maps one tensor block to a CUDA thread block, thus the balanced workload by non-zero distribution for COO-\MTTKRP disappears, while the atomic operation stays.
Therefore, the work imbalance due to different numbers of non-zeros in tensor blocks could be make its performance even worse than COO-\MTTKRP-GPU.

\section{Tensor Dataset} \label{sec:dataset} 
\begin{table}[htbp]
\centering
\scriptsize
{
\caption{Description of sparse tensors.}
\label{tab:real-tensors}
\begin{tabular}{rr|cccc}
  \textbf{No.} & \textbf{Tensors} & \textbf{Order} & \textbf{Dimensions} & \textbf{\#Nnzs} & \textbf{Density} \\
  \toprule
  r1 & \tennm{vast} & 3 & $165K \times 11K \times 2$ & 26M & $6.9 \times 10^{-3}$ \\
  r2 & \tennm{nell2} & 3 & $12K \times 9K \times 29K$ & 77M & $2.4 \times 10^{-5}$ \\
  r3 & \tennm{choa} & 3 & $712K \times 10K \times 767$ & 27M & $5.0 \times 10^{-6}$ \\
  r4 & \tennm{darpa} & 3 & $22K \times 22K \times 24M$ & 28M & $2.4 \times 10^{-9}$ \\
  r5 & \tennm{fb-m} & 3 & $23M \times 23M \times 166$ & 100M & $1.1 \times 10^{-9}$ \\
  r6 & \tennm{fb-s} & 3 & $39M \times 39M \times 532$ & 140M & $1.7 \times 10^{-10}$ \\
  r7 & \tennm{flickr} & 3 & $320K \times 28M \times 1.6M$ & 113M & $7.8 \times 10^{-12}$ \\
  r8 & \tennm{deli} & 3 & $533K \times 17M \times 2.5M$ & 140M & $6.1 \times 10^{-12}$ \\
  r9 & \tennm{nell1} & 3 & $2.9M \times 2.1M \times 25M$ & 144M & $9.1 \times 10^{-13}$ \\ 
  \midrule
  r10 & \tennm{crime4d} & 4 & $6K \times 24 \times 77 \times 32$ & 5M & $1.5 \times 10^{-2}$ \\
  r11 & \tennm{uber4d} & 4 & $183 \times 24 \times 1140 \times 1717$ & 3M & $3.9 \times 10^{-4}$ \\
  r12 & \tennm{nips4d} & 4 & $2K \times 3K \times 14K \times 17$ & 3M & $1.8 \times 10^{-6}$ \\ 
  r13 & \tennm{enron4d} & 4 & $6K \times 6K \times 244K \times 1K$ & 54M & $5.5 \times 10^{-9}$ \\ 
  r14 & \tennm{flickr4d} & 4 & $320K \times 28M \times 1.6M \times 731$ & 113M & $1.1 \times 10^{-14}$ \\ 
  r15 & \tennm{deli4d} & 4 & $533K \times 17M \times 2.5M \times 1K$ & 140M & $4.3 \times 10^{-15}$ \\ 
  \bottomrule
\end{tabular}
}
\end{table}
 
This benchmark suite uses sparse tensors derived from real-world applications from online collections~\cite{Smith:2017:frostt-dataset,Jeon:2015:haten2,Perros:2017:spartan}. It also generates synthetic sparse tensors based on graph models that preserve the properties of real-world graphs. 
\footnote{Our benchmark suite includes synthetic tensor generators, and links to the tensors already included in existed collections~\cite{Smith:2017:frostt-dataset,Jeon:2015:haten2,Perros:2017:spartan}.}
The benchmark suite can be run against any set of tensors provided that they are expressed using coordinate format.

\subsection{Tensors From Real World Applications}\label{subsection:realTensors}
The tensors taken from real-world applications are described in Table~\ref{tab:real-tensors} sorted by tensor order and decreasing density.
They are taken from The Formidable Repository of Open Sparse Tensors and Tools (FROSTT) dataset~\cite{Smith:2017:frostt-dataset}, the HaTen2~\cite{Jeon:2015:haten2-pak} dataset, and one built from electronic medical records from Children's Healthcare of Atlanta~\cite{Perros:2017:spartan}.
\rv{These tensors were chosen to represent a wide range of domains: pattern recognition (\tennm{vast}, \tennm{nips4d}), natural language processing (\tennm{nell2}, \tennm{nell1}), healthcare analytics (\tennm{choa}), recommendation systems (\tennm{deli}, \tennm{deli4d}, \tennm{flickr4d}), crime detection (\tennm{crime4d}), anomaly detection (\tennm{enron4d}).}

%

\subsection{Synthetic Tensor Generation}

The Kronecker graph model~\cite{Leskovec:2010:KroneckerGraphs} and the biased power law generator from the FireHose streaming benchmark~\cite{Anderson:2015:Firehose} were extended to generate 3- and 4-dimensional tensors.
These methods were selected because the resultant graphs follow the power law distribution, exhibit a small diameter, and have a high average clustering coefficient.

The software to generate synthetic tensors is included in the benchmark suite and the synthetic tensors used in our experiments are described in \Cref{tab:syn-tensors} in a period order of "small, medium, large". The regular 3D and 4D tensors, which are equidimensional along each mode, are generated using the Kronecker generator. The irregular 3D and 4D tensors are generated using the biased power law  streaming generator. The 3D irregular tensors have one mode completely dense and much smaller compared to the two other modes which are equidimensional  and sparse. Two modes of the irregular 4D tensors are hypersparse and equidimensional, and the other two dimensions are entirely dense and smaller.

\subsubsection{The Stochastic Kronecker Graph Model}

The Stochastic Kronecker graph model~\cite{Leskovec:2010:KroneckerGraphs} is a fractal growth model that preserves the previously listed properties of real-world graphs.
We extended this approach to generate sparse tensors of order $N$ by accepting the initiator as a tensor $\tau_1$ with $N$ modes. 
Similar to the Stochastic Kronecker graph model, by taking the repeated Kronecker product of $\tau _1$ for $n$ times,  a larger $N$th-order tensor $\tau _n$ is produced. 
With Bernoulli sampling, $\tau _n$ can be realized as a large sparse tensor representing the resultant hyper-graph that follow the properties of real-world networks. 

The exponential growth of Kronecker multiplication limits the sizes of $N$th-order tensors that can be generated. 
We overcome this by performing an additional iteration of Kronecker multiplication and strip off the tensor coordinates that fall outside the given size along each mode. 

\subsubsection{The Power Law Generator} 
The power law generated graphs do not possess a high average clustering coefficient, there is no restriction  on the size of the graph generated.
The generator produces a stream of edges that when combined form a graph respecting the power law distribution.
This is used to create tensors by combining together the sparse graphs to form slices of a third order tensor which is hypergraph. 
This process, when repeated on 3rd order tensors can generate a sparse tensor with $N$ modes.

\begin{table}[tb]
\centering
\scriptsize
\caption{Description of synthetic tensors generated (Gen.) with Kronecker and Power law generators indicated as Kron. and PL respectively.}
\label{tab:syn-tensors}
\begin{tabular}{rr|cp{0.3cm}ccc}
\textbf{No.} & \textbf{Tensors} & \textbf{Gen.} & \textbf{Order} & \textbf{Dimensions} & \textbf{\#Nnzs} & \textbf{Density} \\
\toprule
s1 & \tennm{regS} & Kron. & 3 & $(65K)^3$ & 1.1M & $3.72 \times 10^{-9}$ \\
s2 & \tennm{regM} & Kron. & 3 & $(1.1M)^3$ & 11.5M  & $9.97 \times 10^{-12}$ \\
s3 & \tennm{regL} & Kron. & 3 & $(8.3M)^3$  & 94M   & $1.61 \times 10^{-13}$  \\
s4 & \tennm{irrS} & PL & 3 & $(32K)^2 \times 76 $ & 1M & $1.26 \times 10^{-5}$ \\
s5 & \tennm{irrM} & PL & 3 & $(524K)^2 \times 126$ & 10M & $1.43 \times 10^{-6}$ \\
s6 & \tennm{irrL} & PL & 3 & $(4.2M)^2 \times 168$ & 84M & $2.86 \times 10^{-8}$ \\
\midrule
s7 & \tennm{regS4d} & Kron. & 4 & $(8.2K)^4$ & 1M & $2.26 \times 10^{-10}$ \\
s8 & \tennm{regM4d} & Kron. & 4 & $(2.1M)^4$ & 11.2M & $5.83 \times 10^{-19}$ \\
s9 & \tennm{regL4d} & Kron. & 4 & $(8.3M)^4$ & 110M & $2.23 \times 10^{-20}$ \\
s10 & \tennm{irrS4d} & PL & 4 & $(1.6M)^3 \times 82$ & 1.0M & $2.90 \times 10^{-9}$ \\
s11 & \tennm{irrM4d} & PL & 4 & $(2.6M)^3 \times 144$ & 10.8M & $4.17 \times 10^{-12}$ \\
s12 & \tennm{irrL4d} & PL & 4 & $(4.2M)^3 \times 226$ & 100M  & $6.0\times 10^{-15}$ \\
s13 & \tennm{irr2S4d} & PL & 4 & $(1.0M)^2 \times 122 \times 436$ & 1.6M & $2.81 \times 10^{-11}$ \\
s14 & \tennm{irr2M4d} & PL & 4 & $(4.2M)^2 \times 232 \times 746$ & 19.9M & $6.53 \times 10^{-12}$ \\
s15 & \tennm{irr2L4d} & PL & 4 & $(8.3M)^2 \times 952 \times 324$ & 109M  & $5.1\times 10^{-12}$ \\
\bottomrule
\end{tabular}
\end{table}

\section{Experimental Results} \label{sec:exp}
We test our tensor kernels on four parallel platforms including Intel CPUs and NVIDIA GPUs and build Roofline performance models to measure our performance bounds for detailed analysis. 

\subsection{Configurations}

\subsubsection{Platforms}
We run the experiments on four parallel platforms, the parameters of which are listed in \Cref{tab:platform}.
All Intel platforms are non-uniform memory access (NUMA) machines with 2-4 NUMA nodes.
We calculate the peak single-precision (SP) floating point performance and main/global-memory bandwidth from these parameters.
The peak SP performance of all machines is above 1 TFLOPS. 
GPUs show advantages in peak performance and memory bandwidth over CPUs by approximately $4-12 \times$ and $3-7 \times$ respectively.

\begin{table}[htbp]
\centering
\scriptsize
{
\caption{Platform parameters.}
\label{tab:platform}
\begin{tabular}{rcccc}
  \toprule
  
  \multirow{2}{*}{\textbf{Parameters}} &
  \multicolumn{2}{c}{\textbf{Intel CPUs}} &
  \multicolumn{2}{c}{\textbf{NVIDIA GPUs}} \\ \cmidrule(lr){2-3} \cmidrule(lr){4-5}

   & 
  \textbf{Bluesky} & 
  \textbf{Wingtip} & 
  \textbf{DGX-1P} &
  \textbf{DGX-1V} \\ \midrule

  \multirow{2}{*}{Processor} &
  Intel Xeon &
  Intel Xeon &
  NVIDIA &
  NVIDIA \\

   &
  Gold 6126 &
  E7-4850v3 &
  Tesla P100 &
  Tesla V100 \\

  Microarch &
  Skylake &
  Haswell &
  Pascal &
  Volta \\

  Frequency &
  2.60 GHz &
  2.20 GHz &
  1.48 GHz &
  1.53 GHz \\

  \#Cores &
  24 ($12 \times 2$) &
  56 ($14 \times 4$) &
  3584 &
  5120 \\

  Peak SP &
  1.0 &
  2.0 &
  10.6 &
  14.9 \\

  Perf. &
  TFLOPS &
  TFLOPS &
  TFLOPS &
  TFLOPS \\

  LLC size &
  19 MB &
  35 MB &
  3 MB &
  6 MB \\

  Mem. size &
  196 GB &
  2114 GB &
  16 GB &
  16 GB \\

  Mem. type &
  DDR4 &
  DDR4 &
  HBM2 &
  HBM2 \\

  Mem. freq. &
  2.666 GHz &
  2.133 GHz &
  0.715 GHz &
  0.877 GHz \\

  Mem. BW &
  256 GB/s &
  273 GB/s &
  732 GB/s &
  900 GB/s \\

  Compiler &
  gcc 7.1.0 &
  gcc 5.5.0 &
  CUDA Tkit 9.1 &
  CUDA Tkit 9.0\\

  \bottomrule
\end{tabular}
}
\end{table}

\subsubsection{Kernel Implementation Details}
Since our data is naturally sorted in a particular mode order, COO implementations could take some advantages from the better data locality.
The addition operation is the representative of \TEW and the multiplication operation represents \TS; the performance using different operations are quite similar in our experiments. 
For multicore CPU implementations, we use OpenMP for parallelization with different scheduling strategies, with the number of threads is set to the number of physical cores.
``omp atomic'' is used to deal with data race in \MTTKRP, and ``omp simd'' is for vectorization of \TTM and \MTTKRP.
We use ``numactl --interleave=all --physcpubind'' to interleave memory allocation for better memory bandwidth usage and thread binding for lower scheduling overhead. 
For GPU implementations, ``atomicAdd'' is used in \MTTKRP.
For HiCOO format, we fix the block size to $128$ to fit into the last-level cache in all platforms and use only 8 bits to store element indices.
We use $16$ as the column size for matrices in \TTM and \MTTKRP, to reflect the low-rank feature in popular tensor methods.
We run all kernels five times to get the average; the time of \TTV, \TTM, and \MTTKRP is further averaged among all tensor modes.


\begin{figure}[tbh]
  \centering
  \footnotesize
  \begin{tabular}{cc}
  \includegraphics[width=0.46\linewidth]{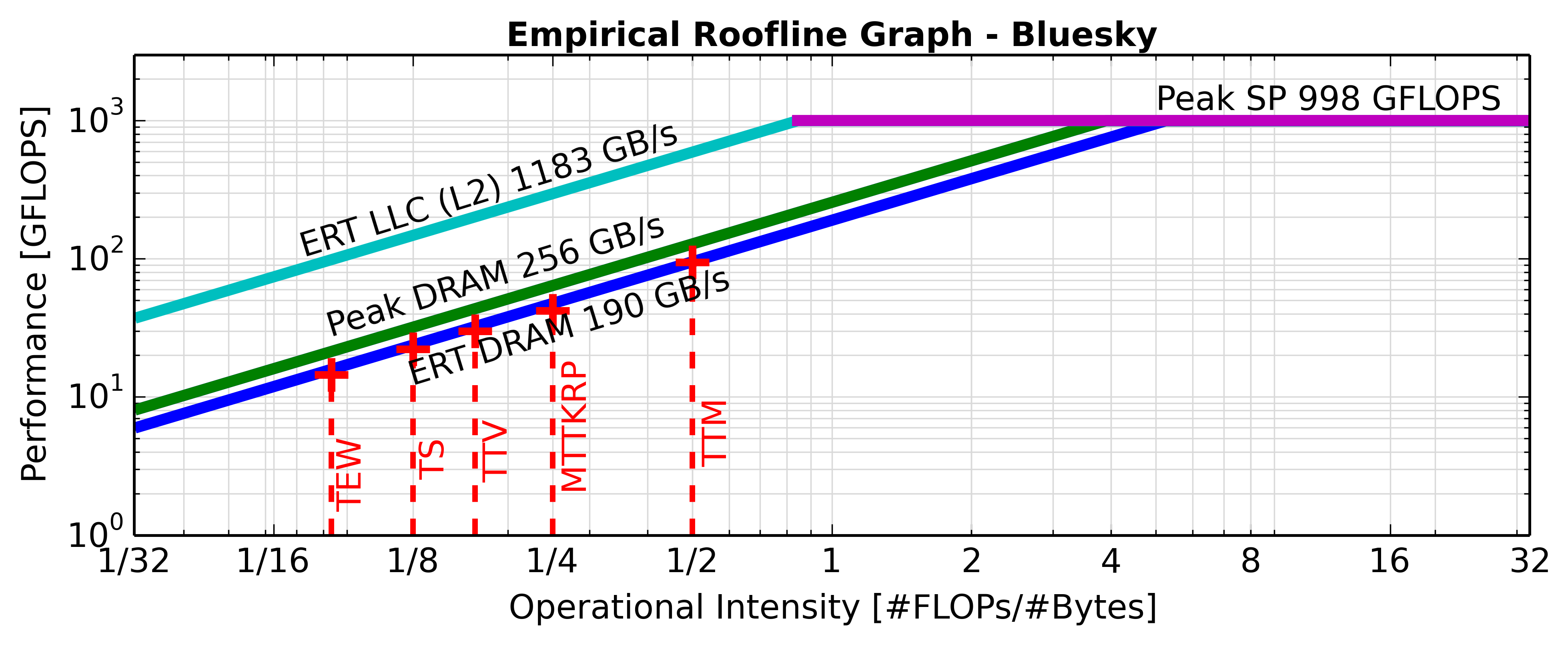} &
  \includegraphics[width=0.46\linewidth]{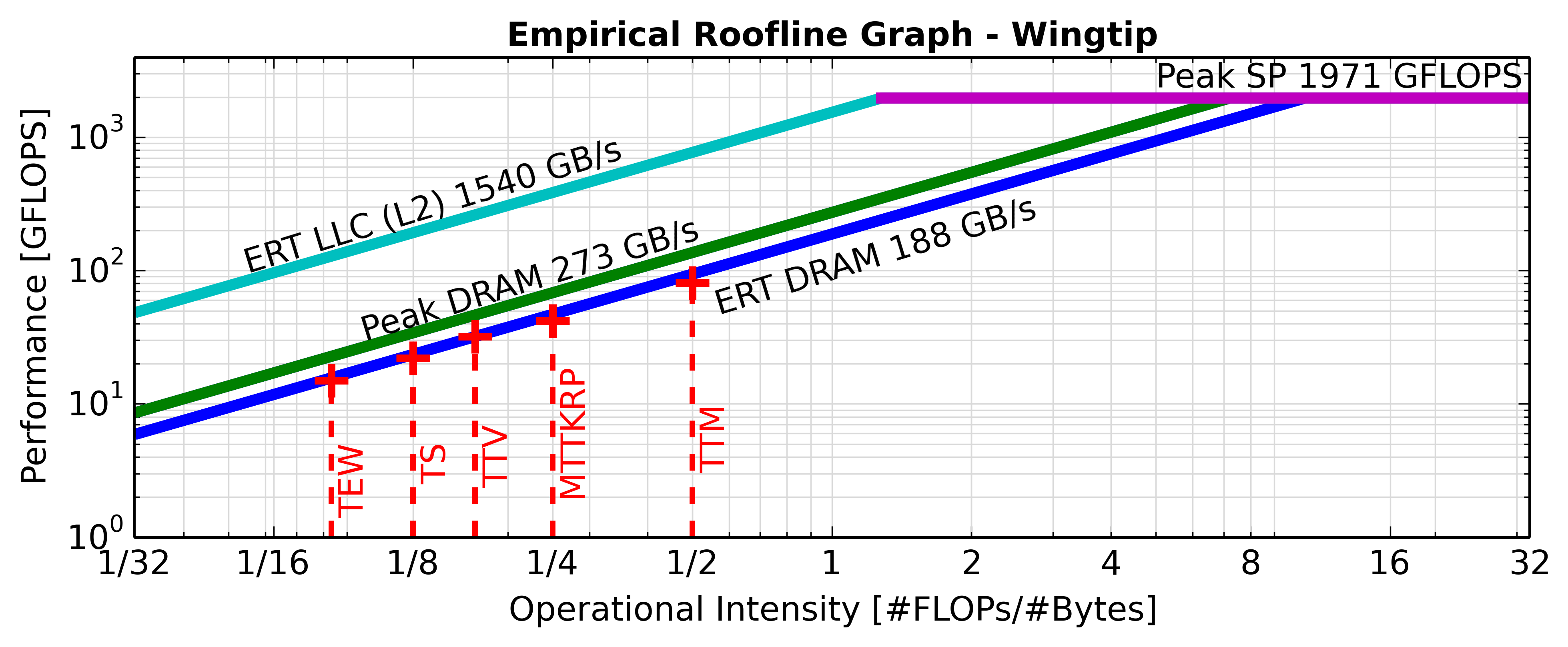} \\
  (a) Bluesky &
  (b) Wingtip \\

  \includegraphics[width=0.46\linewidth]{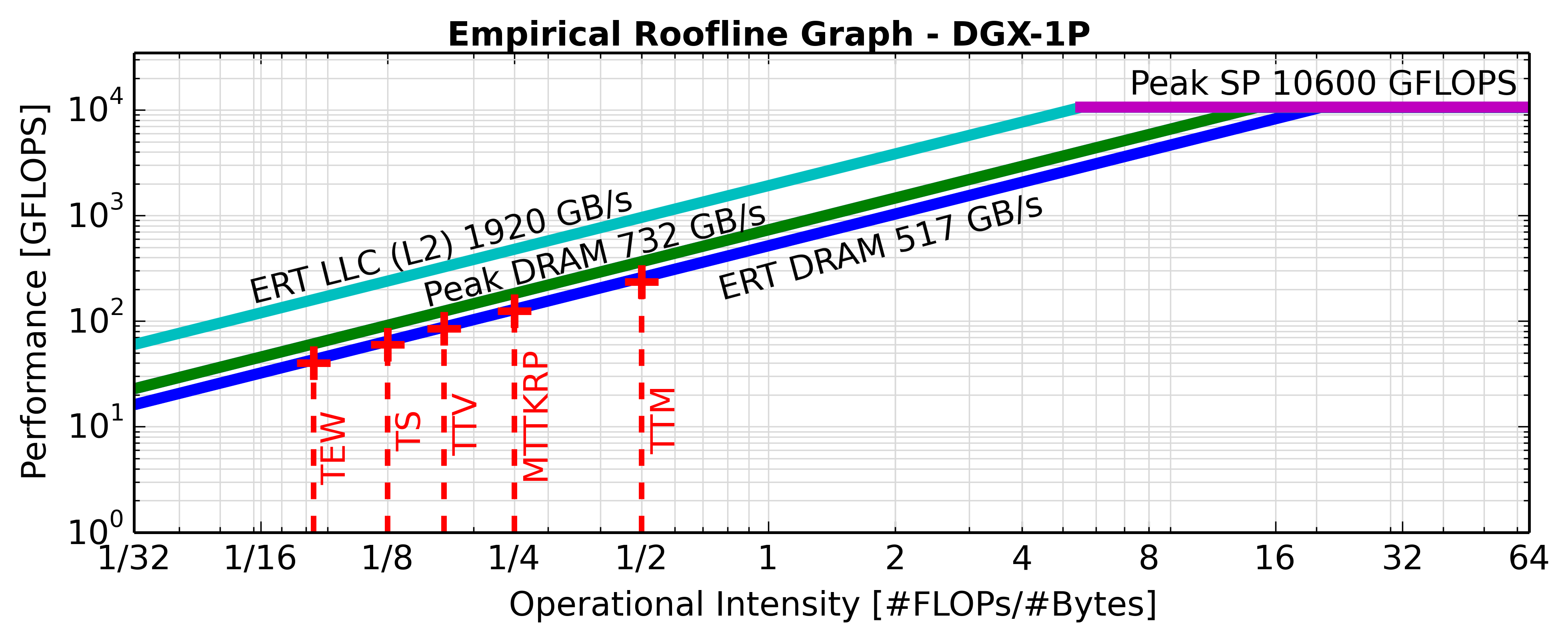} &
  \includegraphics[width=0.46\linewidth]{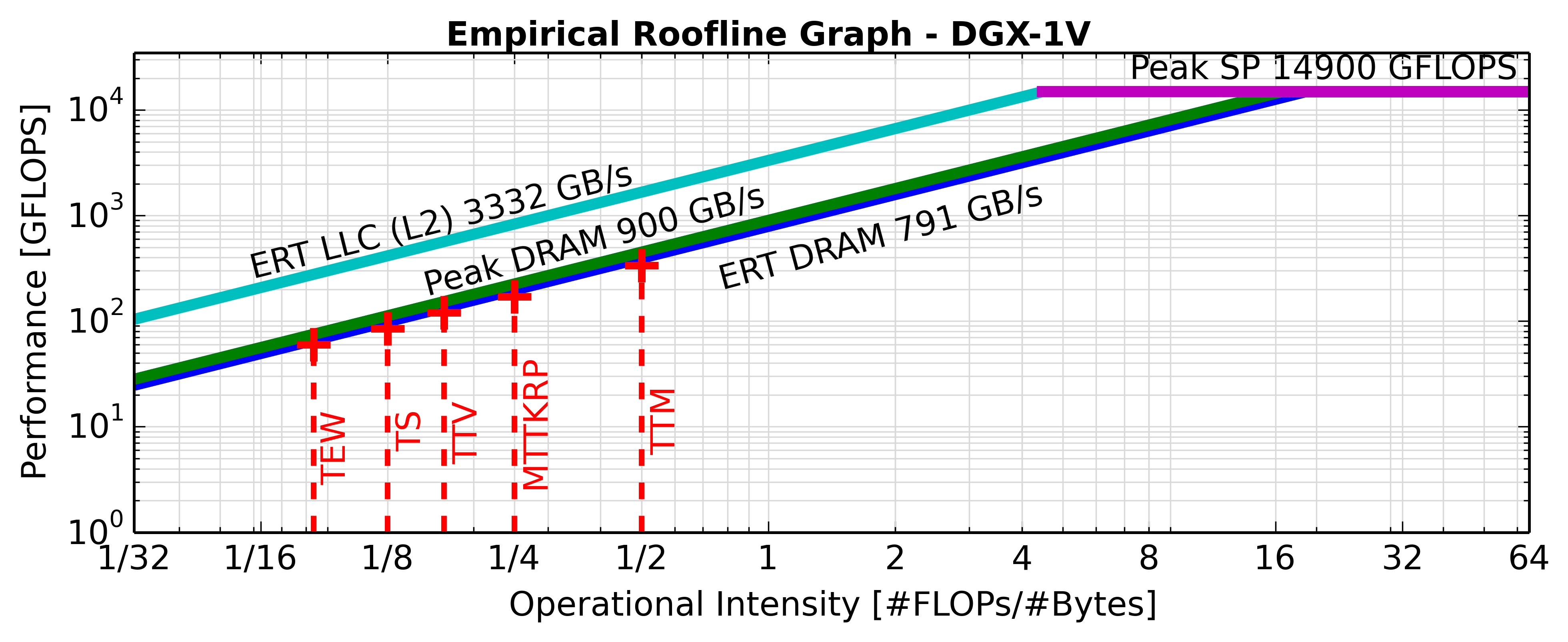} \\
  (c) DGX-1P &
  (d) DGX-1V \\
  \end{tabular}
  \caption{Roofline models marked with the operational intensities of tensor kernels.}
  \label{fig:roofline}
\end{figure}

\subsection{Roofline Performance Models}

The Roofline performance model~\cite{Williams:2009:roofline,Williams:2008:thesis} is a graphical representation of machine characteristics.
It is employed for performance analysis in various application domains: digital signal processing, e.g., Spiral ~\cite{Ofenbeck:14}, sparse/dense linear algebra~\cite{Williams:2008:thesis,Zhang:2017}, and Lattice Boltzmann Magnetohydrodynamics (LBMHD)~\cite{Williams:2008:thesis}.
The Empirical Roofline Tool (ERT)~\cite{Lo:2015:ERT}, included into Intel Advisor tool, automates measuring the target machine characteristics.
ERT automatically generates Roofline data including the maximum bandwidth for various levels of the memory hierarchy which are obtained by testing a variety of micro-kernels \footnote{similar to STREAM benchmark suite~\cite{mccalpin1995stream}}.
ERT can utilize MPI, OpenMP, and CUDA for parallelization; we configure it to the corresponding compiler in \Cref{tab:platform} for tests.

\Cref{fig:roofline} plots the Roofline models for the four platforms in \Cref{tab:platform} with DRAM and last-level cache (LLC) bandwidth tested from ERT, and the theoretical peak SP performance and DRAM bandwidth (not cache-aware) for reference.
We mark the operational intensities (\#Flops/\#Bytes) of our tensor kernels calculated from \Cref{tab:analysis} overlying with Roofline models.
``ERT-DRAM'' bandwidth is the obtainable bandwidth from benchmarking micro-kernels, thus OIs of all the tensor kernels are marked on this line.
From this figure, all of the sparse tensor kernels we consider are main or global memory bound for CPUs and GPUs respectively.
Higher bandwidth will accelerate kernel execution, while other factors such as better data reuse (cache utilization) lowering memory access pressure will also lead to performance improvement.
We use the computed obtainable performance of all tensor kernels as the upper bounds in our performance figures below (called ``Roofline performance''), calculated by timing an OI value with the ``ERT-DRAM'' bandwidth. 
The OI value is an accurate \#Flops/\#Bytes ratio by taking different tensor features into account, especially for \TTV and \TTM because of the $\nfibs$ term in \Cref{tab:analysis}.

\begin{figure*}[h]
  \centering
  \footnotesize
  \includegraphics[width=0.85\linewidth]{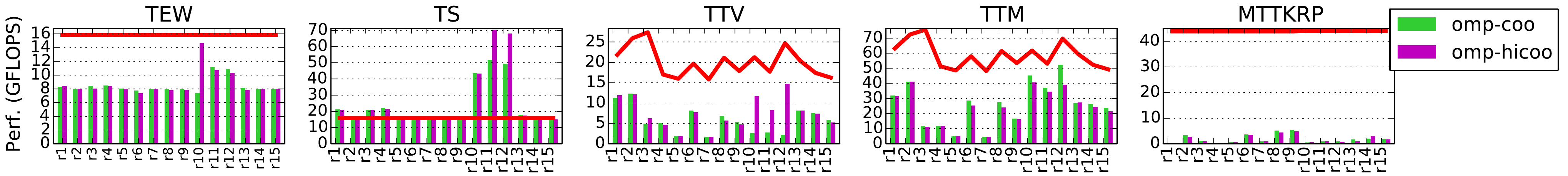} \\
  (a) Real tensors \\
  \includegraphics[width=0.85\linewidth]{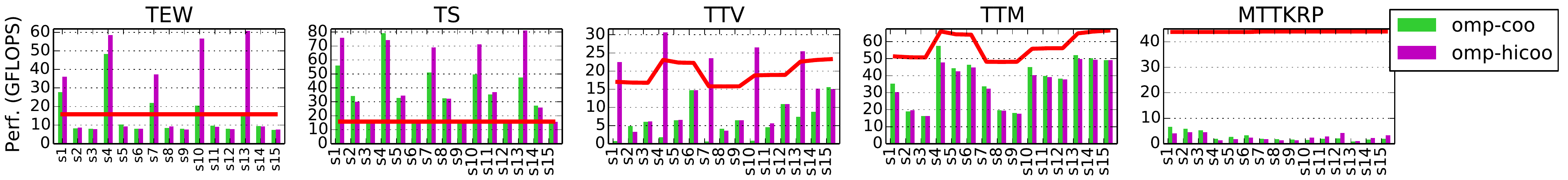} \\
  (b) Synthetic tensors \\
  \caption{Single-precision performance of tensor kernels on the Bluesky platform with the Roofline performance.}
  \label{fig:bluesky}
\end{figure*}


\begin{figure*}[h]
  \centering
  \footnotesize
  \includegraphics[width=0.85\linewidth]{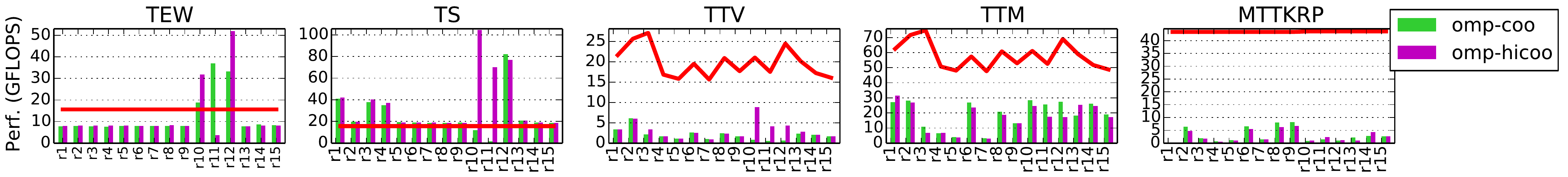} \\
  (a) Real tensors \\
  \includegraphics[width=0.85\linewidth]{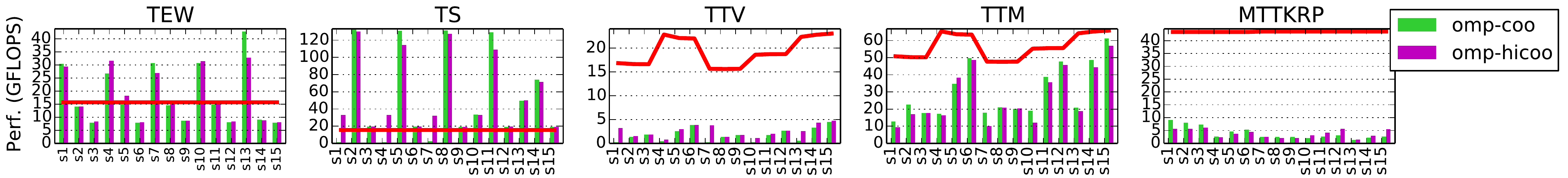} \\
  (b) Synthetic tensors \\
  \caption{Single-precision performance of tensor kernels on the Wingtip platform with the Roofline performance.}
  \label{fig:wingtip}
\end{figure*}

\begin{figure*}[h]
  \centering
  \footnotesize
  \includegraphics[width=0.85\linewidth]{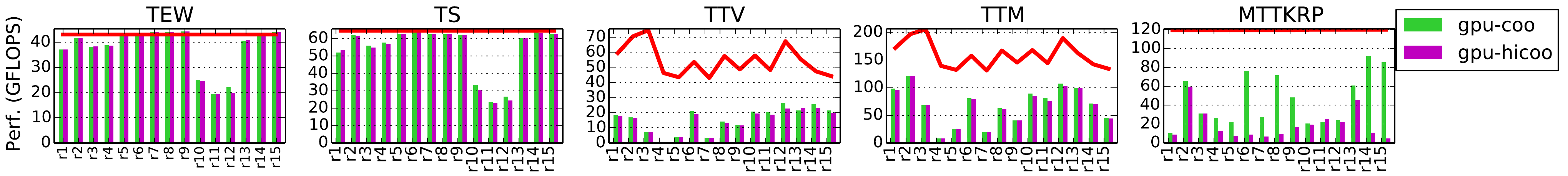} \\
  (a) Real tensors \\
  \includegraphics[width=0.85\linewidth]{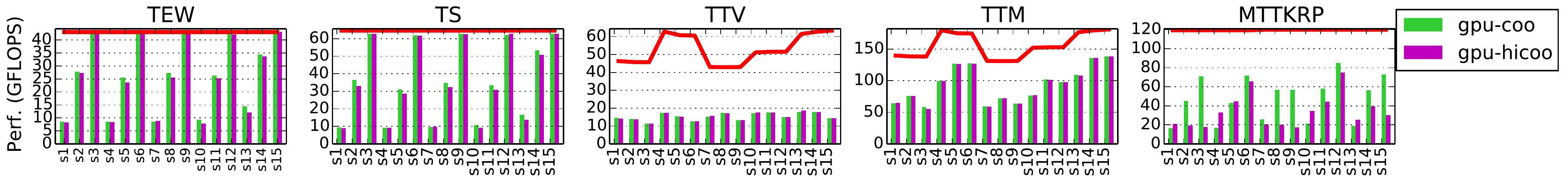} \\

  (b) Synthetic tensors \\
  \caption{Single-precision performance of tensor kernels on NVIDIA DGX-1P with the Roofline performance.}
  \label{fig:p100}
\end{figure*}

\begin{figure*}[h]
  \centering
  \footnotesize
  \includegraphics[width=0.85\linewidth]{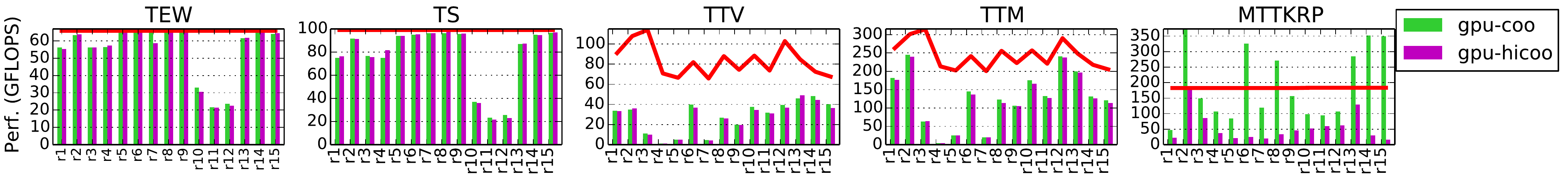} \\
  (a) Real tensors \\
  \includegraphics[width=0.85\linewidth]{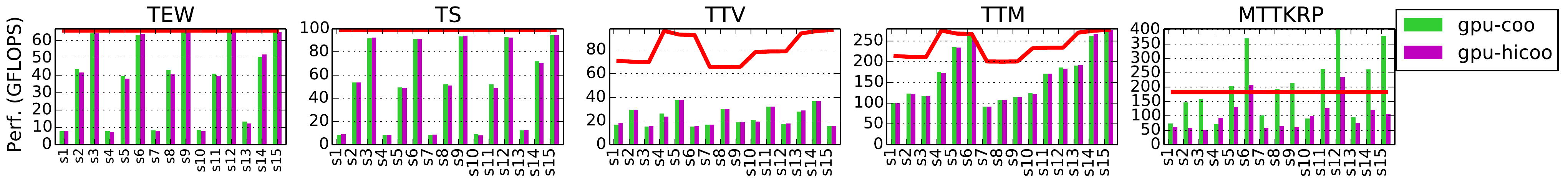} \\
  (b) Synthetic tensors \\
  \caption{Single-precision performance of tensor kernels on NVIDIA DGX-1V with the Roofline performance.}
  \label{fig:v100}
\end{figure*}


\subsection{Performance}

This section presents the performance of all the five tensor kernels on two datasets, real and synthetic, for four platforms.
The performance in GFLOPS of each tensor kernel is calculated from \#Flops (in \Cref{tab:analysis}) divided by the measured execution time.
X-axis represents tensors using the numbers in \Cref{tab:real-tensors,tab:syn-tensors} from different datasets.


\textbf{Observation 1:} \emph{Achieved performance is diverse and hard to predict, which varies with the dimension sizes and non-zero patterns of tensors, platforms, and data formats.} 

From \Cref{fig:bluesky,fig:wingtip,fig:p100,fig:v100}, the actual performance in GFLOPS are extremely diverse between tensor kernels, data formats, platforms, and datasets.
Take synthetic dataset as an example, the achieved performance varies from 0.8 GFLOPS (\tennm{regS} in COO) to 81 GLOPS (\tennm{irr2S4d} in HiCOO) in \Cref{fig:bluesky} on Bluesky platform.
Besides, the average performance of the five kernels ranges a lot as well. 
\TEW, \TS, \TTV, \TTM, \MTTKRP kernels achieve an average of $14.6$, $35.1$, $6.3$, $37.7$, and $2.7$ GFLOPS respectively for COO format, and $22.3$, $40.8$, $14.4$, $35.8$, and $2.6$ GFLOPS respectively for HiCOO format.
This also shows HiCOO on average behaves better than COO format for \TEW, \TS, and \TTV and gets similar performance on \TTM and \MTTKRP.
Even for the performance efficiency (or bandwidth efficiency), these kernels still vary a large range from the lowest 2\% (\MTTKRP on \tennm{irr2S4d}) to 353\% (\TS on \tennm{regS}) for COO format and 2\% (\MTTKRP on \tennm{irr2S4d}) - 479\% (\TS on \tennm{regS}). (The above 100\% efficiency phenomenon will be explained below.)
Also, quite different performance numbers are observed between real and synthetic datasets under the same tensor kernel.
For Wingtip platform in \Cref{fig:wingtip}, \TTV shows much lower GFLOPS numbers than those on Bluesky.
Though we observe some trend especially for synthetic dataset and \TTM operation, generally it is hard to predict the performance of a sparse kernel, even operating with a dense matrix or vector.

\textbf{Observation 2:} \emph{Performance is generally below the Roofline performance calculated from main/global memory bandwidth except for some small tensors fitting into caches or algorithms with good data locality thus making a good use of caches.} 

Most of cases in \Cref{fig:bluesky,fig:wingtip,fig:p100,fig:v100} fall below the red Roofline performance line calculated from main/global memory bandwidth from \Cref{fig:roofline} except for some case of \TEW and \TS on Bluesky and Wingtip CPU platforms and \MTTKRP on DGX-1V GPU platform.
Take \Cref{fig:bluesky} as an example again, the tensors exceed Roofline performance are all small tensors with around 1M non-zeros: \tennm{regS}, \tennm{irrS}, \tennm{regS4d}, \tennm{irrS4d}, and \tennm{irr2S4d} for \TEW, all small and medium synthetic tensors and small real tensors with 3-5M non-zeros: \tennm{crime4d}, \tennm{uber4d}, and \tennm{nips4d}.
The last level cache size of Bluesky is 19 MB which could reside three tensor values each with around 1.6M non-zeros of \TEW and two tensor values each with around 2.4M non-zeros of \TS.
These numbers matches with the small synthetic and real tensors, while the medium tensors also gain some cache benefit.
For \MTTKRP on DGX-1V, COO-\MTTKRP-GPU gets higher performance than Roofline more on irregular-shaped tensors in synthetic dataset. 
And the tensors achieve high performance on DGX-1P are easier to get break the upper bound on DGX-1V.
One reason is that V100 GPU architecture has a twice larger LLC (6M) than P100; besides, V100 get improved atomic operation performance which could benefit \MTTKRP; another reason is that they may have very good data reuse or small working-set size, e.g., tensors with a very short mode, so it is cache that offers the data injection rate rather than off-chip memory; lastly, the integer and floating-point operations have independent data-paths for instruction issuing on Volta architecture. Therefore, address computation which is extensively used in \MTTKRP can be overlapped with floating-point operations, which may mitigate the waiting time for address calculation compared with earlier GPU architectures.

\textbf{Observation 3:} \emph{It is hard to obtain good performance efficiency for non-streaming kernels on multi-socket CPU machines because of NUMA effect, which could be even harder than on GPUs.} 

On Bluesky (\Cref{fig:bluesky}), the average performance efficiency of \TTV, \TTM, and \MTTKRP is 31\%, 64\%, 6\% for COO format, and 73\%, 61\%, 5\% for HiCOO format; while the numbers are 9\%, 52\%, 9\% for COO and 13\%, 47\%, 9\% for HiCOO on Wingtip (\Cref{fig:wingtip}).
Though \MTTKRP behaves a little higher efficiency on Wingtip, its efficiency is still very low. The increment could come from better parallelism of Wingtip with 56 cores.
On DGX-1P GPU (\Cref{fig:p100}), the average performance efficiency of \TTV, \TTM, and \MTTKRP is 30\%, 60\%, 40\% for COO format, and 30\%, 60\%, 28\% for HiCOO format; while the numbers are 30\%, 69\%, 110\% for COO and 30\%, 69\%, 57\% for HiCOO on DGX-1V (\Cref{fig:v100}).
The efficiency numbers on four-socket Wingtip CPU are all lower than those on DGX-1P and DGX-1V GPUs, while \TTV and \TTM achieves better or similar efficiency than the two GPUs on two-socket Bluesky CPU.

\textbf{Observation 4:} \emph{HiCOO algorithms is faster than or similar to COO counterparts because of its better local locality and smaller memory footprint, except \MTTKRP on GPUs where load imbalance and lower parallelism play more important roles.} 

From the average efficiency and performance numbers shown in Observation 1 and 3, HiCOO on average behaves better than COO format for \TEW, \TS, and \TTV and gets similar performance on \TTM and \MTTKRP on CPU platforms. 
On the two GPUs, due to their smaller last-level cache size, HiCOO does not benefit as much as on CPUs. 
From \Cref{fig:p100,fig:v100}, HiCOO gets very similar performance on \TEW, \TS, \TTV, and \TTM because of their similar execution code for tensor value computation. 
While HiCOO-\MTTKRP behaves worse than COO-\MTTKRP because of their different parallel strategies.
HiCOO parallelize tensor blocks with severe load imbalance issue and lower parallelism compared to COO-\MTTKRP algorithm.
Thus, to better use HiCOO format, a careful tuning need to be done according to architecture features.

\textbf{Observation 5:} \emph{Different datasets expose very different performance behavior, which shows the importance of synthetic datasets to performance benchmarking and analysis.} 

Compare the performance trend of real and synthetic datasets. 
\TEW and \TS show obvious period trend from high to low or low to high on CPUs and GPUs respectively due to different cache sizes on synthetic dataset, while it is hard to find trends in real dataset.
\TTV and especially \TTM show a matching trend with the Roofline predicted performance for both real and synthetic tensors.
Since real tensors are from diverse real application scenarios, it is hard to do benchmarking and performance analysis solely based on them. 
Besides, real tensors are limited due to data privacy and other publicity issues. 
Extracting features from real tensors as a basis to create more complete synthetic tensors would be very helpful for sparse tensor research.



Overall, diverse performance behavior is observed among the five kernels and between the COO and HiCOO formats on the Intel CPU and NVIDIA GPU platforms.
Our benchmark performance is still lower than the theoretical Roofline for most cases, especially for \MTTKRP.
Advanced performance optimization could be further explored, such as the lock-avoiding techniques employed in~\cite{Li:2018:hicoo}.

\section{Related Work} \label{sec:rel}

Work related to this benchmark suite includes various benchmarking collections and synthetic benchmark data generation.
This benchmark suite is distinct from previous efforts in its focus on multi-dimensional tensors combined with both real and synthetic input.

Benchmark suites measure machine attributes and exemplify computing patterns.
Benchmarks that measure specific machine attributes include LINPACK~\cite{dongarra2003linpack}, SPEC~\cite{dixit1991spec}, STREAM~\cite{mccalpin1995stream}, GeekBench~\cite{primate:geekbench}, Multimaps~\cite{snavely2002framework}, Bandwidth~\cite{smith2008bandwidth}, and others.
Most of them aim to measure memory bandwidth achieved under varying conditions and a few target architecture floating-point capabilities.
Benchmark suites are often organized around the concept of application exemplars.
These suites emulate common patterns and behaviors in application classes of interest.
Several examples of these suites have been published:  LAPACK/ScaLAPACK for dense linear algebra~\cite{blackford96}, Colella's Seven Motif's~\cite{colella7} for scientific computing, PARSEC~\cite{bienia2008parsec} and SPLASH2 \cite{woo1995splash}, Rodinia \cite{che2009Rodinia}, Graph500~\cite{murphy2010introducing}, SparseBench~\cite{dongarra2001sparsebench}, GAP~\cite{beamer2015gap}, SSCA\#2~\cite{kepner2005hpcs}, and Tartan~\cite{li2018tartan} are just a few.


Several approaches to synthetic graph generation have been proposed. Our work extends two of these, power law graphs from Firehose, and Kronecker graphs from Graph500~\cite{murphy2010introducing}. 
FireHose is a suite of stream processing benchmarks~\cite{Anderson:2015:Firehose}, one of a front-end generator of which is the biased power law generator.
Existing synthetic tensor generators like SimTensor~\cite{Fanaee:2016:Simtensor}, Nway Toolbox~\cite{Andersson:2000:nway}, and the Tensor Toolbox~\cite{Bader:2017:tensortoolbox-pak} are specific to tensors with Tucker~\cite{Tucker:1966:tucker}, CANDECOMP/PARAFAC decomposition~\cite{Carroll:1970:cp,Carroll:1980:candelinc} structures or particular data distributions.
This paper provides a starting point to generate sparse tensors that preserve the properties of real-world or multi-attributed graphs that can be realized as higher-order sparse tensors.

\rv{Many libraries support sparse tensor methods, such as Tensor Toolbox~\cite{Bader:2017:tensortoolbox-pak}, Nway Toolbox~\cite{Andersson:2000:nway}, Tensorlab~\cite{Vervliet:2016:tensorlab-pak}, TACO~\cite{kjolstad:2017:taco}, SPLATT~\cite{Smith:2015:splatt}, and ParTI~\cite{Li:2016:parti-pak}. 
As a benchmark suite, we supply widely-adopted reference implementations and make continuously effort to include state-of-the-art algorithms and data structures as well.}



\section{Conclusion} \label{sec:con}
This paper presents a benchmark suite targeting sparse tensor kernels.
Operations on sparse tensors are common in a wide range of important applications.
The operations are memory bound and often dominate application performance.
This benchmark suite identifies important kernels and data representations and provides reference implementations to aid the community in effectively sharing and comparing performance and optimization results.
Two methods for synthetic tensor generation are provided by preserving the properties of real-world graphs. A subset of possible synthetic tensors are used in this paper. The tool provides the ability to generate custom synthetic tensors in a reproducible manner.

Five observations are made based on performance analysis over Roofline models to gain insights of sparse tensor behavior across architectures.
This benchmark suite is a continuous effort: additional operations, such as TTM-chain in Tucker decomposition, tensor contraction, a sparse tensor with a sparse vector/matrix operations; more complete tensor methods, such as CANDECOMP/PARAFAC and Tucker decompositions; data representations, such as compressed sparse fiber (CSF)~\cite{Smith:2015:splatt}, balanced CSF (BCSF)~\cite{Nisa:2019:BCSF}; more platforms, such as distributed systems, multiple GPUs, and other new architectures (e.g., FPGAs and Emu~\cite{Dysart:2016:emu,Hein:2018:emu}) will be included to the suite in the future.




\FloatBarrier

%


%
\bibliographystyle{ACM-Reference-Format}
\bibliography{ppopp20}


\begin{thebibliography}{67}


\ifx \showCODEN    \undefined \def \showCODEN     #1{\unskip}     \fi
\ifx \showDOI      \undefined \def \showDOI       #1{#1}\fi
\ifx \showISBNx    \undefined \def \showISBNx     #1{\unskip}     \fi
\ifx \showISBNxiii \undefined \def \showISBNxiii  #1{\unskip}     \fi
\ifx \showISSN     \undefined \def \showISSN      #1{\unskip}     \fi
\ifx \showLCCN     \undefined \def \showLCCN      #1{\unskip}     \fi
\ifx \shownote     \undefined \def \shownote      #1{#1}          \fi
\ifx \showarticletitle \undefined \def \showarticletitle #1{#1}   \fi
\ifx \showURL      \undefined \def \showURL       {\relax}        \fi
\providecommand\bibfield[2]{#2}
\providecommand\bibinfo[2]{#2}
\providecommand\natexlab[1]{#1}
\providecommand\showeprint[2][]{arXiv:#2}

\bibitem[\protect\citeauthoryear{Anandkumar, Ge, Hsu, Kakade, and
  Telgarsky}{Anandkumar et~al\mbox{.}}{2014}]%
        {Anandkumar:2014:survey}
\bibfield{author}{\bibinfo{person}{Animashree Anandkumar},
  \bibinfo{person}{Rong Ge}, \bibinfo{person}{Daniel Hsu},
  \bibinfo{person}{Sham~M. Kakade}, {and} \bibinfo{person}{Matus Telgarsky}.}
  \bibinfo{year}{2014}\natexlab{}.
\newblock \showarticletitle{Tensor Decompositions for Learning Latent Variable
  Models}.
\newblock \bibinfo{journal}{\emph{J. Mach. Learn. Res.}} \bibinfo{volume}{15},
  \bibinfo{number}{1} (\bibinfo{date}{Jan.} \bibinfo{year}{2014}),
  \bibinfo{pages}{2773--2832}.
\newblock
\showISSN{1532-4435}


\bibitem[\protect\citeauthoryear{Anderson and Plimpton}{Anderson and
  Plimpton}{2015}]%
        {Anderson:2015:Firehose}
\bibfield{author}{\bibinfo{person}{Karl Anderson} {and} \bibinfo{person}{Steve
  Plimpton}.} \bibinfo{year}{2015}\natexlab{}.
\newblock \bibinfo{booktitle}{\emph{FireHose Streaming Benchmarks}}.
\newblock \bibinfo{type}{{T}echnical {R}eport}. \bibinfo{institution}{Sandia
  National Laboratory}.
\newblock


\bibitem[\protect\citeauthoryear{Andersson and Bro}{Andersson and Bro}{2000}]%
        {Andersson:2000:nway}
\bibfield{author}{\bibinfo{person}{Claus~A Andersson} {and}
  \bibinfo{person}{Rasmus Bro}.} \bibinfo{year}{2000}\natexlab{}.
\newblock \showarticletitle{The {N-way} {Toolbox} for {MATLAB}}.
\newblock \bibinfo{journal}{\emph{Chemometrics and Intelligent Laboratory
  Systems}} \bibinfo{volume}{52}, \bibinfo{number}{1} (\bibinfo{date}{Aug}
  \bibinfo{year}{2000}), \bibinfo{pages}{1--4}.
\newblock


\bibitem[\protect\citeauthoryear{Bader, Kolda, et~al\mbox{.}}{Bader
  et~al\mbox{.}}{2017}]%
        {Bader:2017:tensortoolbox-pak}
\bibfield{author}{\bibinfo{person}{Brett~W. Bader}, \bibinfo{person}{Tamara~G.
  Kolda}, {et~al\mbox{.}}} \bibinfo{year}{2017}\natexlab{}.
\newblock \bibinfo{title}{{MATLAB Tensor Toolbox} ({V}ersion 3.0-dev)}.
\newblock \bibinfo{howpublished}{Available online}.
\newblock
\urldef\tempurl%
\url{https://www.tensortoolbox.org}
\showURL{%
\tempurl}


\bibitem[\protect\citeauthoryear{Baskaran, Meister, Vasilache, and
  Lethin}{Baskaran et~al\mbox{.}}{2012}]%
        {Baskaran:2012:sparse-tensor}
\bibfield{author}{\bibinfo{person}{M. Baskaran}, \bibinfo{person}{B. Meister},
  \bibinfo{person}{N. Vasilache}, {and} \bibinfo{person}{R. Lethin}.}
  \bibinfo{year}{2012}\natexlab{}.
\newblock \showarticletitle{Efficient and scalable computations with sparse
  tensors}. In \bibinfo{booktitle}{\emph{High Performance Extreme Computing
  (HPEC), 2012 IEEE Conference on}}. \bibinfo{pages}{1--6}.
\newblock
\urldef\tempurl%
\url{https://doi.org/10.1109/HPEC.2012.6408676}
\showDOI{\tempurl}


\bibitem[\protect\citeauthoryear{Beamer, Asanovi{\'c}, and Patterson}{Beamer
  et~al\mbox{.}}{2015}]%
        {beamer2015gap}
\bibfield{author}{\bibinfo{person}{Scott Beamer}, \bibinfo{person}{Krste
  Asanovi{\'c}}, {and} \bibinfo{person}{David Patterson}.}
  \bibinfo{year}{2015}\natexlab{}.
\newblock \showarticletitle{The GAP benchmark suite}.
\newblock \bibinfo{journal}{\emph{arXiv preprint arXiv:1508.03619}}
  (\bibinfo{year}{2015}).
\newblock


\bibitem[\protect\citeauthoryear{Bienia, Kumar, Singh, and Li}{Bienia
  et~al\mbox{.}}{2008}]%
        {bienia2008parsec}
\bibfield{author}{\bibinfo{person}{Christian Bienia}, \bibinfo{person}{Sanjeev
  Kumar}, \bibinfo{person}{Jaswinder~Pal Singh}, {and} \bibinfo{person}{Kai
  Li}.} \bibinfo{year}{2008}\natexlab{}.
\newblock \showarticletitle{The {PARSEC} benchmark suite: Characterization and
  architectural implications}. In \bibinfo{booktitle}{\emph{Proceedings of the
  17th international conference on Parallel architectures and compilation
  techniques}}. ACM, \bibinfo{pages}{72--81}.
\newblock


\bibitem[\protect\citeauthoryear{Blackford, Choi, Cleary, Petitet, Whaley,
  Demmel, Dhillon, Stanley, Dongarra, Hammarling, Henry, and Walker}{Blackford
  et~al\mbox{.}}{1996}]%
        {blackford96}
\bibfield{author}{\bibinfo{person}{Laura~Susan Blackford}, \bibinfo{person}{J.
  Choi}, \bibinfo{person}{A. Cleary}, \bibinfo{person}{A. Petitet},
  \bibinfo{person}{R.~C. Whaley}, \bibinfo{person}{J. Demmel},
  \bibinfo{person}{I. Dhillon}, \bibinfo{person}{K. Stanley},
  \bibinfo{person}{J. Dongarra}, \bibinfo{person}{S. Hammarling},
  \bibinfo{person}{G. Henry}, {and} \bibinfo{person}{D. Walker}.}
  \bibinfo{year}{1996}\natexlab{}.
\newblock \showarticletitle{ScaLAPACK: A Portable Linear Algebra Library for
  Distributed Memory Computers - Design Issues and Performance}. In
  \bibinfo{booktitle}{\emph{Proceedings of the 1996 ACM/IEEE Conference on
  Supercomputing}} \emph{(\bibinfo{series}{Supercomputing '96})}.
  \bibinfo{publisher}{IEEE Computer Society}, \bibinfo{address}{Washington, DC,
  USA}, Article \bibinfo{articleno}{5}.
\newblock
\showISBNx{0-89791-854-1}
\urldef\tempurl%
\url{https://doi.org/10.1145/369028.369038}
\showDOI{\tempurl}


\bibitem[\protect\citeauthoryear{Carroll and Chang}{Carroll and Chang}{1970}]%
        {Carroll:1970:cp}
\bibfield{author}{\bibinfo{person}{J.~Douglas Carroll} {and}
  \bibinfo{person}{Jih-Jie Chang}.} \bibinfo{year}{1970}\natexlab{}.
\newblock \showarticletitle{Analysis of individual differences in
  multidimensional scaling via an n-way generalization of ``Eckart-Young''
  decomposition}.
\newblock \bibinfo{journal}{\emph{Psychometrika}} \bibinfo{volume}{35},
  \bibinfo{number}{3} (\bibinfo{date}{01 Sep} \bibinfo{year}{1970}),
  \bibinfo{pages}{283--319}.
\newblock
\showISSN{1860-0980}
\urldef\tempurl%
\url{https://doi.org/10.1007/BF02310791}
\showDOI{\tempurl}


\bibitem[\protect\citeauthoryear{Carroll, Pruzansky, and Kruskal}{Carroll
  et~al\mbox{.}}{1980}]%
        {Carroll:1980:candelinc}
\bibfield{author}{\bibinfo{person}{J.~D. Carroll}, \bibinfo{person}{S.
  Pruzansky}, {and} \bibinfo{person}{J.~B. Kruskal}.}
  \bibinfo{year}{1980}\natexlab{}.
\newblock \showarticletitle{{C}{A}{N}{D}{E}{L}{I}{N}{C}: {A} general approach
  to multidimensional analysis of many-way arrays with linear constraints on
  parameters}.
\newblock \bibinfo{journal}{\emph{Psychometrika}}  \bibinfo{volume}{45}
  (\bibinfo{year}{1980}), \bibinfo{pages}{3--24}.
\newblock


\bibitem[\protect\citeauthoryear{Che, Boyer, Meng, Tarjan, Sheaffer, Lee, and
  Skadron}{Che et~al\mbox{.}}{2009}]%
        {che2009Rodinia}
\bibfield{author}{\bibinfo{person}{S. Che}, \bibinfo{person}{M. Boyer},
  \bibinfo{person}{J. Meng}, \bibinfo{person}{D. Tarjan},
  \bibinfo{person}{J.~W. Sheaffer}, \bibinfo{person}{S. Lee}, {and}
  \bibinfo{person}{K. Skadron}.} \bibinfo{year}{2009}\natexlab{}.
\newblock \showarticletitle{{Rodinia}: A benchmark suite for heterogeneous
  computing}. In \bibinfo{booktitle}{\emph{2009 IEEE International Symposium on
  Workload Characterization (IISWC)}}. \bibinfo{pages}{44--54}.
\newblock
\urldef\tempurl%
\url{https://doi.org/10.1109/IISWC.2009.5306797}
\showDOI{\tempurl}


\bibitem[\protect\citeauthoryear{Chou, Kjolstad, and Amarasinghe}{Chou
  et~al\mbox{.}}{2018}]%
        {chou:2018:formats}
\bibfield{author}{\bibinfo{person}{Stephen Chou}, \bibinfo{person}{Fredrik
  Kjolstad}, {and} \bibinfo{person}{Saman Amarasinghe}.}
  \bibinfo{year}{2018}\natexlab{}.
\newblock \showarticletitle{Format Abstraction for Sparse Tensor Algebra
  Compilers}.
\newblock \bibinfo{journal}{\emph{Proc. ACM Program. Lang.}}
  \bibinfo{volume}{2}, \bibinfo{number}{OOPSLA}, Article
  \bibinfo{articleno}{123} (\bibinfo{date}{Oct.} \bibinfo{year}{2018}),
  \bibinfo{numpages}{30}~pages.
\newblock
\showISSN{2475-1421}
\urldef\tempurl%
\url{https://doi.org/10.1145/3276493}
\showDOI{\tempurl}


\bibitem[\protect\citeauthoryear{Cichocki}{Cichocki}{2014}]%
        {Cichocki:2014:survey}
\bibfield{author}{\bibinfo{person}{Andrzej Cichocki}.}
  \bibinfo{year}{2014}\natexlab{}.
\newblock \showarticletitle{Era of Big Data Processing: {A} New Approach via
  Tensor Networks and Tensor Decompositions}.
\newblock \bibinfo{journal}{\emph{CoRR}}  \bibinfo{volume}{abs/1403.2048}
  (\bibinfo{year}{2014}).
\newblock


\bibitem[\protect\citeauthoryear{{Cichocki}, {Lee}, {Oseledets}, {Phan},
  {Zhao}, and {Mandic}}{{Cichocki} et~al\mbox{.}}{2016}]%
        {Cichocki:2016:survey}
\bibfield{author}{\bibinfo{person}{A. {Cichocki}}, \bibinfo{person}{N. {Lee}},
  \bibinfo{person}{I.~V. {Oseledets}}, \bibinfo{person}{A. {Phan}},
  \bibinfo{person}{Q. {Zhao}}, {and} \bibinfo{person}{D. {Mandic}}.}
  \bibinfo{year}{2016}\natexlab{}.
\newblock \showarticletitle{Low-Rank Tensor Networks for Dimensionality
  Reduction and Large-Scale Optimization Problems: Perspectives and Challenges
  PART 1}.
\newblock \bibinfo{journal}{\emph{ArXiv e-prints}} (\bibinfo{date}{Sept.}
  \bibinfo{year}{2016}).
\newblock
\showeprint[arxiv]{cs.NA/1609.00893}


\bibitem[\protect\citeauthoryear{Colella}{Colella}{2004}]%
        {colella7}
\bibfield{author}{\bibinfo{person}{Phil Colella}.}
  \bibinfo{year}{2004}\natexlab{}.
\newblock \showarticletitle{Defining Software Requirements for Scientific
  Computing}.
\newblock  (\bibinfo{date}{01} \bibinfo{year}{2004}).
\newblock


\bibitem[\protect\citeauthoryear{De~Lathauwer, Vervliet, Boussé, and
  Debals}{De~Lathauwer et~al\mbox{.}}{2017}]%
        {DeLathauwer:2017:tensorize}
\bibfield{author}{\bibinfo{person}{Lieven De~Lathauwer}, \bibinfo{person}{Nico
  Vervliet}, \bibinfo{person}{Martijn Boussé}, {and} \bibinfo{person}{Otto
  Debals}.} \bibinfo{year}{2017}\natexlab{}.
\newblock \bibinfo{title}{Dealing with curse and blessing of dimensionality
  through tensor decompositions}.
\newblock
\newblock


\bibitem[\protect\citeauthoryear{{Di Napoli}, Fabregat-Traver, Quintana-Ortí,
  and Bientinesi}{{Di Napoli} et~al\mbox{.}}{2014}]%
        {DiNapoli:2014:tc}
\bibfield{author}{\bibinfo{person}{Edoardo {Di Napoli}}, \bibinfo{person}{Diego
  Fabregat-Traver}, \bibinfo{person}{Gregorio Quintana-Ortí}, {and}
  \bibinfo{person}{Paolo Bientinesi}.} \bibinfo{year}{2014}\natexlab{}.
\newblock \showarticletitle{Towards an efficient use of the {BLAS} library for
  multilinear tensor contractions}.
\newblock \bibinfo{journal}{\emph{Appl. Math. Comput.}}  \bibinfo{volume}{235}
  (\bibinfo{year}{2014}), \bibinfo{pages}{454 -- 468}.
\newblock
\showISSN{0096-3003}
\urldef\tempurl%
\url{https://doi.org/10.1016/j.amc.2014.02.051}
\showDOI{\tempurl}


\bibitem[\protect\citeauthoryear{Dixit}{Dixit}{1991}]%
        {dixit1991spec}
\bibfield{author}{\bibinfo{person}{Kaivalya~M Dixit}.}
  \bibinfo{year}{1991}\natexlab{}.
\newblock \showarticletitle{The {SPEC} benchmarks}.
\newblock \bibinfo{journal}{\emph{Parallel computing}} \bibinfo{volume}{17},
  \bibinfo{number}{10-11} (\bibinfo{year}{1991}), \bibinfo{pages}{1195--1209}.
\newblock


\bibitem[\protect\citeauthoryear{Dongarra, Eijkhout, and van~der
  Vorst}{Dongarra et~al\mbox{.}}{2001}]%
        {dongarra2001sparsebench}
\bibfield{author}{\bibinfo{person}{Jack Dongarra}, \bibinfo{person}{Victor
  Eijkhout}, {and} \bibinfo{person}{Henk van~der Vorst}.}
  \bibinfo{year}{2001}\natexlab{}.
\newblock \bibinfo{title}{SparseBench: A sparse iterative benchmark}.
\newblock
\newblock


\bibitem[\protect\citeauthoryear{Dongarra, Luszczek, and Petitet}{Dongarra
  et~al\mbox{.}}{2003}]%
        {dongarra2003linpack}
\bibfield{author}{\bibinfo{person}{Jack~J Dongarra}, \bibinfo{person}{Piotr
  Luszczek}, {and} \bibinfo{person}{Antoine Petitet}.}
  \bibinfo{year}{2003}\natexlab{}.
\newblock \showarticletitle{The LINPACK benchmark: past, present and future}.
\newblock \bibinfo{journal}{\emph{Concurrency and Computation: practice and
  experience}} \bibinfo{volume}{15}, \bibinfo{number}{9}
  (\bibinfo{year}{2003}), \bibinfo{pages}{803--820}.
\newblock


\bibitem[\protect\citeauthoryear{Dysart, Kogge, Deneroff, Bovell, Briggs,
  Brockman, Jacobsen, Juan, Kuntz, Lethin, McMahon, Pawar, Perrigo, Rucker,
  Ruttenberg, Ruttenberg, and Stein}{Dysart et~al\mbox{.}}{2016}]%
        {Dysart:2016:emu}
\bibfield{author}{\bibinfo{person}{Timothy Dysart}, \bibinfo{person}{Peter
  Kogge}, \bibinfo{person}{Martin Deneroff}, \bibinfo{person}{Eric Bovell},
  \bibinfo{person}{Preston Briggs}, \bibinfo{person}{Jay Brockman},
  \bibinfo{person}{Kenneth Jacobsen}, \bibinfo{person}{Yujen Juan},
  \bibinfo{person}{Shannon Kuntz}, \bibinfo{person}{Richard Lethin},
  \bibinfo{person}{Janice McMahon}, \bibinfo{person}{Chandra Pawar},
  \bibinfo{person}{Martin Perrigo}, \bibinfo{person}{Sarah Rucker},
  \bibinfo{person}{John Ruttenberg}, \bibinfo{person}{Max Ruttenberg}, {and}
  \bibinfo{person}{Steve Stein}.} \bibinfo{year}{2016}\natexlab{}.
\newblock \showarticletitle{Highly Scalable Near Memory Processing with
  Migrating Threads on the {Emu} System Architecture}. In
  \bibinfo{booktitle}{\emph{Proceedings of the Sixth Workshop on Irregular
  Applications: Architectures and Algorithms}} \emph{(\bibinfo{series}{IA\^3
  '16})}. \bibinfo{publisher}{IEEE Press}, \bibinfo{address}{Piscataway, NJ,
  USA}, \bibinfo{pages}{2--9}.
\newblock
\showISBNx{978-1-5090-3867-1}
\urldef\tempurl%
\url{https://doi.org/10.1109/IA3.2016.7}
\showDOI{\tempurl}


\bibitem[\protect\citeauthoryear{Fanaee-T and Gama}{Fanaee-T and Gama}{2016}]%
        {Fanaee:2016:Simtensor}
\bibfield{author}{\bibinfo{person}{Hadi Fanaee-T} {and} \bibinfo{person}{Joao
  Gama}.} \bibinfo{year}{2016}\natexlab{}.
\newblock \showarticletitle{SimTensor: A synthetic tensor data generator}.
\newblock \bibinfo{journal}{\emph{arXiv preprint arXiv:1612.03772}}
  (\bibinfo{year}{2016}).
\newblock


\bibitem[\protect\citeauthoryear{GeekBench}{GeekBench}{4}]%
        {primate:geekbench}
\bibfield{author}{\bibinfo{person}{GeekBench}.} \bibinfo{year}{4}\natexlab{}.
\newblock \bibinfo{title}{Primate Labs}.
\newblock
\newblock
\urldef\tempurl%
\url{http://http://www.geekbench.com/}
\showURL{%
\tempurl}


\bibitem[\protect\citeauthoryear{Hein, Conte, Young, Eswar, Li, Lavin, Vuduc,
  and Riedy}{Hein et~al\mbox{.}}{2018}]%
        {Hein:2018:emu}
\bibfield{author}{\bibinfo{person}{Eric Hein}, \bibinfo{person}{Tom Conte},
  \bibinfo{person}{Jeffrey~S. Young}, \bibinfo{person}{Srinivas Eswar},
  \bibinfo{person}{Jiajia Li}, \bibinfo{person}{Patrick Lavin},
  \bibinfo{person}{Richard Vuduc}, {and} \bibinfo{person}{Jason Riedy}.}
  \bibinfo{year}{2018}\natexlab{}.
\newblock \showarticletitle{An Initial Characterization of the {Emu} {Chick}}.
\newblock \bibinfo{journal}{\emph{2018 IEEE International Parallel and
  Distributed Processing Symposium Workshops}} (\bibinfo{date}{May}
  \bibinfo{year}{2018}), 10.
\newblock


\bibitem[\protect\citeauthoryear{Israt~Nisa}{Israt~Nisa}{2019}]%
        {Nisa:2019:BCSF}
\bibfield{author}{\bibinfo{person}{Aravind Sukumaran-Rajam Richard Vuduc
  P.~Sadayappan Israt~Nisa, Jiajia~Li}.} \bibinfo{year}{2019}\natexlab{}.
\newblock \bibinfo{title}{Load-balanced sparse {MTTKRP} on {GPUs}}.
\newblock \bibinfo{howpublished}{(To be appeared)}.
\newblock


\bibitem[\protect\citeauthoryear{Jeon, Papalexakis, Kang, and Faloutsos}{Jeon
  et~al\mbox{.}}{2015b}]%
        {Jeon:2015:haten2}
\bibfield{author}{\bibinfo{person}{Inah Jeon}, \bibinfo{person}{Evangelos~E.
  Papalexakis}, \bibinfo{person}{U Kang}, {and} \bibinfo{person}{Christos
  Faloutsos}.} \bibinfo{year}{2015}\natexlab{b}.
\newblock \showarticletitle{{HaTen2}: Billion-scale Tensor Decompositions}. In
  \bibinfo{booktitle}{\emph{IEEE International Conference on Data Engineering
  (ICDE)}}.
\newblock


\bibitem[\protect\citeauthoryear{Jeon, Papalexakis, and U~Kang}{Jeon
  et~al\mbox{.}}{2015a}]%
        {Jeon:2015:haten2-pak}
\bibfield{author}{\bibinfo{person}{Inah Jeon}, \bibinfo{person}{Evangelos~E.
  Papalexakis}, {and} \bibinfo{person}{Christos~Faloutsos U~Kang}.}
  \bibinfo{year}{2015}\natexlab{a}.
\newblock \bibinfo{title}{{HaTen2}: Billion-scale Tensor Decompositions
  ({V}ersion 1.0)}.
\newblock \bibinfo{howpublished}{Available from
  \url{http://datalab.snu.ac.kr/haten2/}}.
\newblock


\bibitem[\protect\citeauthoryear{Ji, Xu, Yang, and Yu}{Ji
  et~al\mbox{.}}{2012}]%
        {ji20123d}
\bibfield{author}{\bibinfo{person}{Shuiwang Ji}, \bibinfo{person}{Wei Xu},
  \bibinfo{person}{Ming Yang}, {and} \bibinfo{person}{Kai Yu}.}
  \bibinfo{year}{2012}\natexlab{}.
\newblock \showarticletitle{{3D} convolutional neural networks for human action
  recognition}.
\newblock \bibinfo{journal}{\emph{IEEE transactions on pattern analysis and
  machine intelligence}} \bibinfo{volume}{35}, \bibinfo{number}{1}
  (\bibinfo{year}{2012}), \bibinfo{pages}{221--231}.
\newblock


\bibitem[\protect\citeauthoryear{Kaya and Uçar}{Kaya and Uçar}{2018}]%
        {Kaya:2018:dimtree-sparse-cp}
\bibfield{author}{\bibinfo{person}{O. Kaya} {and} \bibinfo{person}{B. Uçar}.}
  \bibinfo{year}{2018}\natexlab{}.
\newblock \showarticletitle{Parallel {Candecomp/Parafac} Decomposition of
  Sparse Tensors Using Dimension Trees}.
\newblock \bibinfo{journal}{\emph{SIAM Journal on Scientific Computing}}
  \bibinfo{volume}{40}, \bibinfo{number}{1} (\bibinfo{year}{2018}),
  \bibinfo{pages}{C99--C130}.
\newblock
\urldef\tempurl%
\url{https://doi.org/10.1137/16M1102744}
\showDOI{\tempurl}
\showeprint{https://doi.org/10.1137/16M1102744}


\bibitem[\protect\citeauthoryear{Kepner, Koester, et~al\mbox{.}}{Kepner
  et~al\mbox{.}}{2005}]%
        {kepner2005hpcs}
\bibfield{author}{\bibinfo{person}{J Kepner}, \bibinfo{person}{DP Koester},
  {et~al\mbox{.}}} \bibinfo{year}{2005}\natexlab{}.
\newblock \bibinfo{title}{HPCS SSCA\# 2 Graph Analysis Benchmark Specifications
  v1. 0}.
\newblock
\newblock


\bibitem[\protect\citeauthoryear{Kjolstad, Kamil, Chou, Lugato, and
  Amarasinghe}{Kjolstad et~al\mbox{.}}{2017}]%
        {kjolstad:2017:taco}
\bibfield{author}{\bibinfo{person}{Fredrik Kjolstad}, \bibinfo{person}{Shoaib
  Kamil}, \bibinfo{person}{Stephen Chou}, \bibinfo{person}{David Lugato}, {and}
  \bibinfo{person}{Saman Amarasinghe}.} \bibinfo{year}{2017}\natexlab{}.
\newblock \showarticletitle{The Tensor Algebra Compiler}.
\newblock \bibinfo{journal}{\emph{Proc. ACM Program. Lang.}}
  \bibinfo{volume}{1}, \bibinfo{number}{OOPSLA}, Article
  \bibinfo{articleno}{77} (\bibinfo{date}{Oct.} \bibinfo{year}{2017}),
  \bibinfo{numpages}{29}~pages.
\newblock
\showISSN{2475-1421}
\urldef\tempurl%
\url{https://doi.org/10.1145/3133901}
\showDOI{\tempurl}


\bibitem[\protect\citeauthoryear{Kolda and Bader}{Kolda and Bader}{2009}]%
        {Kolda:2009:survey}
\bibfield{author}{\bibinfo{person}{T. Kolda} {and} \bibinfo{person}{B. Bader}.}
  \bibinfo{year}{2009}\natexlab{}.
\newblock \showarticletitle{Tensor Decompositions and Applications}.
\newblock \bibinfo{journal}{\emph{SIAM Rev.}} \bibinfo{volume}{51},
  \bibinfo{number}{3} (\bibinfo{year}{2009}), \bibinfo{pages}{455--500}.
\newblock
\urldef\tempurl%
\url{https://doi.org/10.1137/07070111X}
\showDOI{\tempurl}
\showeprint{http://dx.doi.org/10.1137/07070111X}


\bibitem[\protect\citeauthoryear{Lebedev, Ganin, Rakhuba, Oseledets, and
  Lempitsky}{Lebedev et~al\mbox{.}}{2014}]%
        {Lebedev:2014:cnn-cp}
\bibfield{author}{\bibinfo{person}{Vadim Lebedev}, \bibinfo{person}{Yaroslav
  Ganin}, \bibinfo{person}{Maksim Rakhuba}, \bibinfo{person}{Ivan Oseledets},
  {and} \bibinfo{person}{Victor Lempitsky}.} \bibinfo{year}{2014}\natexlab{}.
\newblock \showarticletitle{Speeding-up Convolutional Neural Networks Using
  Fine-tuned {CP}-Decomposition}.
\newblock \bibinfo{journal}{\emph{arXiv preprint arXiv:1412.6553}}
  (\bibinfo{year}{2014}).
\newblock


\bibitem[\protect\citeauthoryear{Leskovec, Chakrabarti, Kleinberg, Faloutsos,
  and Ghahramani}{Leskovec et~al\mbox{.}}{2010}]%
        {Leskovec:2010:KroneckerGraphs}
\bibfield{author}{\bibinfo{person}{Jure Leskovec}, \bibinfo{person}{Deepayan
  Chakrabarti}, \bibinfo{person}{Jon Kleinberg}, \bibinfo{person}{Christos
  Faloutsos}, {and} \bibinfo{person}{Zoubin Ghahramani}.}
  \bibinfo{year}{2010}\natexlab{}.
\newblock \showarticletitle{Kronecker graphs: An approach to modeling
  networks}.
\newblock \bibinfo{journal}{\emph{Journal of Machine Learning Research}}
  \bibinfo{volume}{11}, \bibinfo{number}{Feb} (\bibinfo{year}{2010}),
  \bibinfo{pages}{985--1042}.
\newblock


\bibitem[\protect\citeauthoryear{Li, Song, Chen, Liu, Tallent, and Barker}{Li
  et~al\mbox{.}}{2018b}]%
        {li2018tartan}
\bibfield{author}{\bibinfo{person}{Ang Li}, \bibinfo{person}{Shuaiwen~Leon
  Song}, \bibinfo{person}{Jieyang Chen}, \bibinfo{person}{Xu Liu},
  \bibinfo{person}{Nathan Tallent}, {and} \bibinfo{person}{Kevin Barker}.}
  \bibinfo{year}{2018}\natexlab{b}.
\newblock \showarticletitle{{Tartan}: Evaluating Modern {GPU} Interconnect via
  a Multi-{GPU} Benchmark Suite}. In \bibinfo{booktitle}{\emph{2018 IEEE
  International Symposium on Workload Characterization (IISWC)}}. IEEE,
  \bibinfo{pages}{191--202}.
\newblock


\bibitem[\protect\citeauthoryear{Li}{Li}{2018}]%
        {Li:2018:thesis}
\bibfield{author}{\bibinfo{person}{Jiajia Li}.}
  \bibinfo{year}{2018}\natexlab{}.
\newblock \emph{\bibinfo{title}{Scalable tensor decompositions in high
  performance computing environments}}.
\newblock \bibinfo{thesistype}{Ph.D. Dissertation}. \bibinfo{school}{Georgia
  Institute of Technology}, \bibinfo{address}{Atlanta, GA, USA}.
\newblock


\bibitem[\protect\citeauthoryear{Li, Battaglino, Perros, Sun, and Vuduc}{Li
  et~al\mbox{.}}{2015}]%
        {Li:2015:intensli}
\bibfield{author}{\bibinfo{person}{Jiajia Li}, \bibinfo{person}{Casey
  Battaglino}, \bibinfo{person}{Ioakeim Perros}, \bibinfo{person}{Jimeng Sun},
  {and} \bibinfo{person}{Richard Vuduc}.} \bibinfo{year}{2015}\natexlab{}.
\newblock \showarticletitle{An input-adaptive and in-place approach to dense
  tensor-times-matrix multiply}. In \bibinfo{booktitle}{\emph{ACM/IEEE
  Supercomputing (SC '15)}}. \bibinfo{publisher}{ACM}, \bibinfo{address}{New
  York, NY, USA}.
\newblock


\bibitem[\protect\citeauthoryear{Li, Choi, Perros, Sun, and Vuduc}{Li
  et~al\mbox{.}}{2017}]%
        {Li:2017:adatm}
\bibfield{author}{\bibinfo{person}{J. Li}, \bibinfo{person}{J. Choi},
  \bibinfo{person}{I. Perros}, \bibinfo{person}{J. Sun}, {and}
  \bibinfo{person}{R. Vuduc}.} \bibinfo{year}{2017}\natexlab{}.
\newblock \showarticletitle{Model-Driven Sparse {CP} Decomposition for
  Higher-Order Tensors}. In \bibinfo{booktitle}{\emph{2017 IEEE International
  Parallel and Distributed Processing Symposium (IPDPS)}}.
  \bibinfo{pages}{1048--1057}.
\newblock
\urldef\tempurl%
\url{https://doi.org/10.1109/IPDPS.2017.80}
\showDOI{\tempurl}


\bibitem[\protect\citeauthoryear{Li, Ma, and Vuduc}{Li et~al\mbox{.}}{2018a}]%
        {Li:2016:parti-pak}
\bibfield{author}{\bibinfo{person}{Jiajia Li}, \bibinfo{person}{Yuchen Ma},
  {and} \bibinfo{person}{Richard Vuduc}.} \bibinfo{year}{2018}\natexlab{a}.
\newblock \bibinfo{title}{{ParTI!} : A Parallel Tensor Infrastructure for
  multicore {CPUs} and {GPUs} ({V}ersion 1.0.0)}.
\newblock
\newblock
\urldef\tempurl%
\url{https://github.com/hpcgarage/ParTI}
\showURL{%
\tempurl}


\bibitem[\protect\citeauthoryear{Li, Ma, Wu, Li, and Barker}{Li
  et~al\mbox{.}}{2019a}]%
        {li2019pasta}
\bibfield{author}{\bibinfo{person}{Jiajia Li}, \bibinfo{person}{Yuchen Ma},
  \bibinfo{person}{Xiaolong Wu}, \bibinfo{person}{Ang Li}, {and}
  \bibinfo{person}{Kevin Barker}.} \bibinfo{year}{2019}\natexlab{a}.
\newblock \showarticletitle{PASTA: A Parallel Sparse Tensor Algorithm Benchmark
  Suite}.
\newblock \bibinfo{journal}{\emph{arXiv preprint arXiv:1902.03317}}
  (\bibinfo{year}{2019}).
\newblock


\bibitem[\protect\citeauthoryear{Li, Ma, Yan, and Vuduc}{Li
  et~al\mbox{.}}{2016}]%
        {Li:2016:spttm}
\bibfield{author}{\bibinfo{person}{Jiajia Li}, \bibinfo{person}{Yuchen Ma},
  \bibinfo{person}{Chenggang Yan}, {and} \bibinfo{person}{Richard Vuduc}.}
  \bibinfo{year}{2016}\natexlab{}.
\newblock \showarticletitle{Optimizing Sparse Tensor Times Matrix on Multi-core
  and Many-core Architectures}. In \bibinfo{booktitle}{\emph{Proceedings of the
  Sixth Workshop on Irregular Applications: Architectures and Algorithms}}
  \emph{(\bibinfo{series}{IA\^{}3 '16})}. \bibinfo{publisher}{IEEE Press},
  \bibinfo{address}{Piscataway, NJ, USA}, \bibinfo{pages}{26--33}.
\newblock
\showISBNx{978-1-5090-3867-1}
\urldef\tempurl%
\url{https://doi.org/10.1109/IA3.2016.10}
\showDOI{\tempurl}


\bibitem[\protect\citeauthoryear{Li, Sun, and Vuduc}{Li et~al\mbox{.}}{2018c}]%
        {Li:2018:hicoo}
\bibfield{author}{\bibinfo{person}{Jiajia Li}, \bibinfo{person}{Jimeng Sun},
  {and} \bibinfo{person}{Richard Vuduc}.} \bibinfo{year}{2018}\natexlab{c}.
\newblock \showarticletitle{{HiCOO}: {Hierarchical} storage of sparse tensors}.
  In \bibinfo{booktitle}{\emph{Proceedings of the ACM/IEEE International
  Conference on High Performance Computing, Networking, Storage and Analysis
  (SC)}}. \bibinfo{address}{Dallas, TX, USA}.
\newblock


\bibitem[\protect\citeauthoryear{Li, Tan, Chen, and Sun}{Li
  et~al\mbox{.}}{2013}]%
        {Li:2013:smat}
\bibfield{author}{\bibinfo{person}{Jiajia Li}, \bibinfo{person}{Guangming Tan},
  \bibinfo{person}{Mingyu Chen}, {and} \bibinfo{person}{Ninghui Sun}.}
  \bibinfo{year}{2013}\natexlab{}.
\newblock \showarticletitle{{SMAT}: An Input Adaptive Auto-tuner for Sparse
  Matrix-vector Multiplication}. In \bibinfo{booktitle}{\emph{Proceedings of
  the 34th ACM SIGPLAN Conference on Programming Language Design and
  Implementation}} \emph{(\bibinfo{series}{PLDI '13})}.
  \bibinfo{publisher}{ACM}, \bibinfo{address}{New York, NY, USA},
  \bibinfo{pages}{117--126}.
\newblock
\showISBNx{978-1-4503-2014-6}
\urldef\tempurl%
\url{https://doi.org/10.1145/2491956.2462181}
\showDOI{\tempurl}


\bibitem[\protect\citeauthoryear{Li, U\c{c}ar, \c{C}ataly\"{u}rek, Sun, Barker,
  and Vuduc}{Li et~al\mbox{.}}{2019b}]%
        {Li:2019:reorder}
\bibfield{author}{\bibinfo{person}{Jiajia Li}, \bibinfo{person}{Bora U\c{c}ar},
  \bibinfo{person}{\"{U}mit~V. \c{C}ataly\"{u}rek}, \bibinfo{person}{Jimeng
  Sun}, \bibinfo{person}{Kevin Barker}, {and} \bibinfo{person}{Richard Vuduc}.}
  \bibinfo{year}{2019}\natexlab{b}.
\newblock \showarticletitle{Efficient and Effective Sparse Tensor Reordering}.
  In \bibinfo{booktitle}{\emph{Proceedings of the ACM International Conference
  on Supercomputing}} \emph{(\bibinfo{series}{ICS '19})}.
  \bibinfo{publisher}{ACM}, \bibinfo{address}{New York, NY, USA},
  \bibinfo{pages}{227--237}.
\newblock
\showISBNx{978-1-4503-6079-1}
\urldef\tempurl%
\url{https://doi.org/10.1145/3330345.3330366}
\showDOI{\tempurl}


\bibitem[\protect\citeauthoryear{Liu, Wen, Sarwate, and Dehnavi}{Liu
  et~al\mbox{.}}{2017}]%
        {Liu:2017:fcoo}
\bibfield{author}{\bibinfo{person}{B. Liu}, \bibinfo{person}{C. Wen},
  \bibinfo{person}{A.~D. Sarwate}, {and} \bibinfo{person}{M.~M. Dehnavi}.}
  \bibinfo{year}{2017}\natexlab{}.
\newblock \showarticletitle{A Unified Optimization Approach for Sparse Tensor
  Operations on {GPUs}}. In \bibinfo{booktitle}{\emph{2017 IEEE International
  Conference on Cluster Computing (CLUSTER)}}. \bibinfo{pages}{47--57}.
\newblock
\urldef\tempurl%
\url{https://doi.org/10.1109/CLUSTER.2017.75}
\showDOI{\tempurl}


\bibitem[\protect\citeauthoryear{Lo, Williams, Van~Straalen, Ligocki, Cordery,
  Wright, Hall, and Oliker}{Lo et~al\mbox{.}}{2015}]%
        {Lo:2015:ERT}
\bibfield{author}{\bibinfo{person}{Yu~Jung Lo}, \bibinfo{person}{Samuel
  Williams}, \bibinfo{person}{Brian Van~Straalen}, \bibinfo{person}{Terry~J.
  Ligocki}, \bibinfo{person}{Matthew~J. Cordery}, \bibinfo{person}{Nicholas~J.
  Wright}, \bibinfo{person}{Mary~W. Hall}, {and} \bibinfo{person}{Leonid
  Oliker}.} \bibinfo{year}{2015}\natexlab{}.
\newblock \showarticletitle{{Roofline Model Toolkit}: A Practical Tool for
  Architectural and Program Analysis}. In \bibinfo{booktitle}{\emph{High
  Performance Computing Systems. Performance Modeling, Benchmarking, and
  Simulation}}, \bibfield{editor}{\bibinfo{person}{Stephen~A. Jarvis},
  \bibinfo{person}{Steven~A. Wright}, {and} \bibinfo{person}{Simon~D. Hammond}}
  (Eds.). \bibinfo{publisher}{Springer International Publishing},
  \bibinfo{address}{Cham}, \bibinfo{pages}{129--148}.
\newblock
\showISBNx{978-3-319-17248-4}


\bibitem[\protect\citeauthoryear{Ma, Li, Wu, Yan, Sun, and Vuduc}{Ma
  et~al\mbox{.}}{2018}]%
        {Ma:2018:sptucker-gpu}
\bibfield{author}{\bibinfo{person}{Yuchen Ma}, \bibinfo{person}{Jiajia Li},
  \bibinfo{person}{Xiaolong Wu}, \bibinfo{person}{Chenggang Yan},
  \bibinfo{person}{Jimeng Sun}, {and} \bibinfo{person}{Richard Vuduc}.}
  \bibinfo{year}{2018}\natexlab{}.
\newblock \showarticletitle{Optimizing sparse tensor times matrix on {GPUs}}.
\newblock \bibinfo{journal}{\emph{J. Parallel and Distrib. Comput.}}
  (\bibinfo{year}{2018}).
\newblock
\showISSN{0743-7315}
\urldef\tempurl%
\url{https://doi.org/10.1016/j.jpdc.2018.07.018}
\showDOI{\tempurl}


\bibitem[\protect\citeauthoryear{Matthews}{Matthews}{2016}]%
        {Matthews:2016:tc}
\bibfield{author}{\bibinfo{person}{Devin Matthews}.}
  \bibinfo{year}{2016}\natexlab{}.
\newblock \showarticletitle{High-Performance Tensor Contraction without
  {BLAS}}.
\newblock \bibinfo{journal}{\emph{CoRR}}  \bibinfo{volume}{abs/1607.00291}
  (\bibinfo{year}{2016}).
\newblock
\showeprint[arxiv]{1607.00291}
\urldef\tempurl%
\url{http://arxiv.org/abs/1607.00291}
\showURL{%
\tempurl}


\bibitem[\protect\citeauthoryear{McCalpin}{McCalpin}{2007}]%
        {mccalpin1995stream}
\bibfield{author}{\bibinfo{person}{John~D. McCalpin}.}
  \bibinfo{year}{1991-2007}\natexlab{}.
\newblock \showarticletitle{STREAM: Sustainable Memory Bandwidth in High
  Performance Computers}.
\newblock  (\bibinfo{year}{1991-2007}).
\newblock
\urldef\tempurl%
\url{http://www.cs.virginia.edu/stream/}
\showURL{%
\tempurl}
\newblock
\shownote{A continually updated technical report.
  http://www.cs.virginia.edu/stream/.}


\bibitem[\protect\citeauthoryear{Murphy, Wheeler, Barrett, and Ang}{Murphy
  et~al\mbox{.}}{2010}]%
        {murphy2010introducing}
\bibfield{author}{\bibinfo{person}{Richard~C Murphy}, \bibinfo{person}{Kyle~B
  Wheeler}, \bibinfo{person}{Brian~W Barrett}, {and} \bibinfo{person}{James~A
  Ang}.} \bibinfo{year}{2010}\natexlab{}.
\newblock \showarticletitle{Introducing the graph 500}.
\newblock \bibinfo{journal}{\emph{Cray Users Group (CUG)}}
  \bibinfo{volume}{19} (\bibinfo{year}{2010}), \bibinfo{pages}{45--74}.
\newblock


\bibitem[\protect\citeauthoryear{Novikov, Podoprikhin, Osokin, and
  Vetrov}{Novikov et~al\mbox{.}}{2015}]%
        {Novikov:2015:tnn}
\bibfield{author}{\bibinfo{person}{Alexander Novikov}, \bibinfo{person}{Dmitry
  Podoprikhin}, \bibinfo{person}{Anton Osokin}, {and} \bibinfo{person}{Dmitry
  Vetrov}.} \bibinfo{year}{2015}\natexlab{}.
\newblock \showarticletitle{Tensorizing Neural Networks}.
\newblock \bibinfo{journal}{\emph{CoRR}}  \bibinfo{volume}{abs/1509.06569}
  (\bibinfo{year}{2015}).
\newblock


\bibitem[\protect\citeauthoryear{Ofenbeck, Steinmann, Cabezas, Spampinato, and
  P{\"u}schel}{Ofenbeck et~al\mbox{.}}{2014}]%
        {Ofenbeck:14}
\bibfield{author}{\bibinfo{person}{Georg Ofenbeck}, \bibinfo{person}{Ruedi
  Steinmann}, \bibinfo{person}{Victoria~CaparrÃ³s Cabezas},
  \bibinfo{person}{Daniele~G. Spampinato}, {and} \bibinfo{person}{Markus
  P{\"u}schel}.} \bibinfo{year}{2014}\natexlab{}.
\newblock \showarticletitle{Applying the Roofline Model}. In
  \bibinfo{booktitle}{\emph{IEEE International Symposium on Performance
  Analysis of Systems and Software (ISPASS)}}. \bibinfo{pages}{76 -- 85}.
\newblock


\bibitem[\protect\citeauthoryear{Perros, Papalexakis, Wang, Vuduc, Searles,
  Thompson, and Sun}{Perros et~al\mbox{.}}{2017}]%
        {Perros:2017:spartan}
\bibfield{author}{\bibinfo{person}{Ioakeim Perros},
  \bibinfo{person}{Evangelos~E. Papalexakis}, \bibinfo{person}{Fei Wang},
  \bibinfo{person}{Richard Vuduc}, \bibinfo{person}{Elizabeth Searles},
  \bibinfo{person}{Michael Thompson}, {and} \bibinfo{person}{Jimeng Sun}.}
  \bibinfo{year}{2017}\natexlab{}.
\newblock \showarticletitle{{SPARTan}: Scalable {PARAFAC2} for Large \& Sparse
  Data}. In \bibinfo{booktitle}{\emph{Proceedings of the 23rd ACM SIGKDD
  International Conference on Knowledge Discovery and Data Mining}}
  \emph{(\bibinfo{series}{KDD '17})}. \bibinfo{publisher}{ACM},
  \bibinfo{address}{New York, NY, USA}, \bibinfo{pages}{375--384}.
\newblock
\showISBNx{978-1-4503-4887-4}
\urldef\tempurl%
\url{https://doi.org/10.1145/3097983.3098014}
\showDOI{\tempurl}


\bibitem[\protect\citeauthoryear{Sedaghati, Mu, Pouchet, Parthasarathy, and
  Sadayappan}{Sedaghati et~al\mbox{.}}{2015}]%
        {sedaghati2015spmv}
\bibfield{author}{\bibinfo{person}{Naser Sedaghati}, \bibinfo{person}{Te Mu},
  \bibinfo{person}{Louis-Noel Pouchet}, \bibinfo{person}{Srinivasan
  Parthasarathy}, {and} \bibinfo{person}{P. Sadayappan}.}
  \bibinfo{year}{2015}\natexlab{}.
\newblock \showarticletitle{Automatic Selection of Sparse Matrix Representation
  on {GPUs}}. In \bibinfo{booktitle}{\emph{Proceedings of the 29th ACM on
  International Conference on Supercomputing}} \emph{(\bibinfo{series}{ICS
  '15})}. \bibinfo{publisher}{ACM}, \bibinfo{address}{New York, NY, USA},
  \bibinfo{pages}{99--108}.
\newblock
\showISBNx{978-1-4503-3559-1}
\urldef\tempurl%
\url{https://doi.org/10.1145/2751205.2751244}
\showDOI{\tempurl}


\bibitem[\protect\citeauthoryear{Sidiropoulos, {De Lathauwer}, Fu, Huang,
  Papalexakis, and Faloutsos}{Sidiropoulos et~al\mbox{.}}{2017}]%
        {Sidiropoulos:2017:survey}
\bibfield{author}{\bibinfo{person}{N.~D. Sidiropoulos}, \bibinfo{person}{L. {De
  Lathauwer}}, \bibinfo{person}{X. Fu}, \bibinfo{person}{K. Huang},
  \bibinfo{person}{E.~E. Papalexakis}, {and} \bibinfo{person}{C. Faloutsos}.}
  \bibinfo{year}{2017}\natexlab{}.
\newblock \showarticletitle{Tensor Decomposition for Signal Processing and
  Machine Learning}.
\newblock \bibinfo{journal}{\emph{IEEE Transactions on Signal Processing}}
  \bibinfo{volume}{65}, \bibinfo{number}{13} (\bibinfo{date}{July}
  \bibinfo{year}{2017}), \bibinfo{pages}{3551--3582}.
\newblock
\showISSN{1053-587X}
\urldef\tempurl%
\url{https://doi.org/10.1109/TSP.2017.2690524}
\showDOI{\tempurl}


\bibitem[\protect\citeauthoryear{Smith, Choi, Li, Vuduc, Park, Liu, and
  Karypis}{Smith et~al\mbox{.}}{2017}]%
        {Smith:2017:frostt-dataset}
\bibfield{author}{\bibinfo{person}{Shaden Smith}, \bibinfo{person}{Jee~W.
  Choi}, \bibinfo{person}{Jiajia Li}, \bibinfo{person}{Richard Vuduc},
  \bibinfo{person}{Jongsoo Park}, \bibinfo{person}{Xing Liu}, {and}
  \bibinfo{person}{George Karypis}.} \bibinfo{year}{2017}\natexlab{}.
\newblock \bibinfo{title}{{FROSTT}: The {Formidable} {Repository} of {Open}
  {Sparse} {Tensors} and {Tools}}.
\newblock
\newblock
\urldef\tempurl%
\url{http://frostt.io/}
\showURL{%
\tempurl}


\bibitem[\protect\citeauthoryear{Smith, Ravindran, Sidiropoulos, and
  Karypis}{Smith et~al\mbox{.}}{2015}]%
        {Smith:2015:splatt}
\bibfield{author}{\bibinfo{person}{Shaden Smith}, \bibinfo{person}{Niranjay
  Ravindran}, \bibinfo{person}{Nicholas Sidiropoulos}, {and}
  \bibinfo{person}{George Karypis}.} \bibinfo{year}{2015}\natexlab{}.
\newblock \showarticletitle{{SPLATT}: Efficient and Parallel Sparse
  Tensor-Matrix Multiplication}. In \bibinfo{booktitle}{\emph{Proceedings of
  the 29th IEEE International Parallel \& Distributed Processing Symposium}}
  \emph{(\bibinfo{series}{IPDPS})}.
\newblock


\bibitem[\protect\citeauthoryear{Smith}{Smith}{2008}]%
        {smith2008bandwidth}
\bibfield{author}{\bibinfo{person}{Zack Smith}.}
  \bibinfo{year}{2008}\natexlab{}.
\newblock \bibinfo{title}{Bandwidth: a memory bandwidth benchmark}.
\newblock
\newblock


\bibitem[\protect\citeauthoryear{Snavely, Carrington, Wolter, Labarta, Badia,
  and Purkayastha}{Snavely et~al\mbox{.}}{2002}]%
        {snavely2002framework}
\bibfield{author}{\bibinfo{person}{A Snavely}, \bibinfo{person}{L Carrington},
  \bibinfo{person}{N Wolter}, \bibinfo{person}{J Labarta}, \bibinfo{person}{R
  Badia}, {and} \bibinfo{person}{A Purkayastha}.}
  \bibinfo{year}{2002}\natexlab{}.
\newblock \showarticletitle{A framework for performance modeling and
  prediction}. In \bibinfo{booktitle}{\emph{Supercomputing 2002}}. IEEE,
  \bibinfo{pages}{21--21}.
\newblock


\bibitem[\protect\citeauthoryear{Springer, Su, and Bientinesi}{Springer
  et~al\mbox{.}}{2017}]%
        {Springer:2017:HPTT}
\bibfield{author}{\bibinfo{person}{Paul Springer}, \bibinfo{person}{Tong Su},
  {and} \bibinfo{person}{Paolo Bientinesi}.} \bibinfo{year}{2017}\natexlab{}.
\newblock \showarticletitle{{HPTT}: A High-performance Tensor Transposition
  {C++} Library}. In \bibinfo{booktitle}{\emph{Proceedings of the 4th ACM
  SIGPLAN International Workshop on Libraries, Languages, and Compilers for
  Array Programming}} \emph{(\bibinfo{series}{ARRAY 2017})}.
  \bibinfo{publisher}{ACM}, \bibinfo{address}{New York, NY, USA},
  \bibinfo{pages}{56--62}.
\newblock
\showISBNx{978-1-4503-5069-3}
\urldef\tempurl%
\url{https://doi.org/10.1145/3091966.3091968}
\showDOI{\tempurl}


\bibitem[\protect\citeauthoryear{Tucker}{Tucker}{1966}]%
        {Tucker:1966:tucker}
\bibfield{author}{\bibinfo{person}{L.~R. Tucker}.}
  \bibinfo{year}{1966}\natexlab{}.
\newblock \showarticletitle{{S}ome mathematical notes on three-mode factor
  analysis}.
\newblock \bibinfo{journal}{\emph{Psychometrika}}  \bibinfo{volume}{31}
  (\bibinfo{year}{1966}), \bibinfo{pages}{279--311}.
\newblock


\bibitem[\protect\citeauthoryear{{Vervliet}, {Debals}, {Sorber}, {Van Barel},
  and {De Lathauwer}}{{Vervliet} et~al\mbox{.}}{2016}]%
        {Vervliet:2016:tensorlab-pak}
\bibfield{author}{\bibinfo{person}{N. {Vervliet}}, \bibinfo{person}{O.
  {Debals}}, \bibinfo{person}{L. {Sorber}}, \bibinfo{person}{M. {Van Barel}},
  {and} \bibinfo{person}{L. {De Lathauwer}}.} \bibinfo{year}{2016}\natexlab{}.
\newblock \bibinfo{title}{{Tensorlab} ({V}ersion 3.0)}.
\newblock \bibinfo{howpublished}{Available from
  \url{http://www.tensorlab.net}}.
\newblock


\bibitem[\protect\citeauthoryear{Williams, Waterman, and Patterson}{Williams
  et~al\mbox{.}}{2009}]%
        {Williams:2009:roofline}
\bibfield{author}{\bibinfo{person}{Samuel Williams}, \bibinfo{person}{Andrew
  Waterman}, {and} \bibinfo{person}{David Patterson}.}
  \bibinfo{year}{2009}\natexlab{}.
\newblock \showarticletitle{Roofline: An Insightful Visual Performance Model
  for Multicore Architectures}.
\newblock \bibinfo{journal}{\emph{Commun. ACM}} \bibinfo{volume}{52},
  \bibinfo{number}{4} (\bibinfo{date}{April} \bibinfo{year}{2009}),
  \bibinfo{pages}{65--76}.
\newblock
\showISSN{0001-0782}
\urldef\tempurl%
\url{https://doi.org/10.1145/1498765.1498785}
\showDOI{\tempurl}


\bibitem[\protect\citeauthoryear{Williams}{Williams}{2008}]%
        {Williams:2008:thesis}
\bibfield{author}{\bibinfo{person}{Samuel~Webb Williams}.}
  \bibinfo{year}{2008}\natexlab{}.
\newblock \emph{\bibinfo{title}{Auto-tuning Performance on Multicore
  Computers}}.
\newblock \bibinfo{thesistype}{Ph.D. Dissertation}. \bibinfo{school}{EECS
  Department, University of California, Berkeley}.
\newblock
\urldef\tempurl%
\url{http://www.eecs.berkeley.edu/Pubs/TechRpts/2008/EECS-2008-164.html}
\showURL{%
\tempurl}


\bibitem[\protect\citeauthoryear{Woo, Ohara, Torrie, Singh, and Gupta}{Woo
  et~al\mbox{.}}{1995}]%
        {woo1995splash}
\bibfield{author}{\bibinfo{person}{Steven~Cameron Woo},
  \bibinfo{person}{Moriyoshi Ohara}, \bibinfo{person}{Evan Torrie},
  \bibinfo{person}{Jaswinder~Pal Singh}, {and} \bibinfo{person}{Anoop Gupta}.}
  \bibinfo{year}{1995}\natexlab{}.
\newblock \showarticletitle{The SPLASH-2 programs: Characterization and
  methodological considerations}.
\newblock \bibinfo{journal}{\emph{ACM SIGARCH computer architecture news}}
  \bibinfo{volume}{23}, \bibinfo{number}{2} (\bibinfo{year}{1995}),
  \bibinfo{pages}{24--36}.
\newblock


\bibitem[\protect\citeauthoryear{Yu, Zheng, Anandkumar, and Yue}{Yu
  et~al\mbox{.}}{2018}]%
        {Yu:2018:longterm}
\bibfield{author}{\bibinfo{person}{Rose Yu}, \bibinfo{person}{Stephan Zheng},
  \bibinfo{person}{Anima Anandkumar}, {and} \bibinfo{person}{Yisong Yue}.}
  \bibinfo{year}{2018}\natexlab{}.
\newblock \bibinfo{title}{Long-term Forecasting using Tensor-Train {RNN}s}.
\newblock
\newblock
\urldef\tempurl%
\url{https://openreview.net/forum?id=HJJ0w--0W}
\showURL{%
\tempurl}


\bibitem[\protect\citeauthoryear{Zhang, Tan, Xue, Li, Zhou, and Chen}{Zhang
  et~al\mbox{.}}{2017}]%
        {Zhang:2017}
\bibfield{author}{\bibinfo{person}{Xiuxia Zhang}, \bibinfo{person}{Guangming
  Tan}, \bibinfo{person}{Shuangbai Xue}, \bibinfo{person}{Jiajia Li},
  \bibinfo{person}{Keren Zhou}, {and} \bibinfo{person}{Mingyu Chen}.}
  \bibinfo{year}{2017}\natexlab{}.
\newblock \showarticletitle{Understanding the GPU Microarchitecture to Achieve
  Bare-Metal Performance Tuning}. In \bibinfo{booktitle}{\emph{Proceedings of
  the 22Nd ACM SIGPLAN Symposium on Principles and Practice of Parallel
  Programming}} \emph{(\bibinfo{series}{PPoPP '17})}. \bibinfo{publisher}{ACM},
  \bibinfo{address}{New York, NY, USA}, \bibinfo{pages}{31--43}.
\newblock
\showISBNx{978-1-4503-4493-7}
\urldef\tempurl%
\url{https://doi.org/10.1145/3018743.3018755}
\showDOI{\tempurl}


\end{thebibliography}

%

\end{document}